\documentclass{amsart}
\usepackage{graphicx,euscript,eufrak,amsmath,amsthm, color, mathrsfs}

\usepackage[margin=1.0in]{geometry}

\newcommand{\bx}{\boldsymbol{x}}
\newcommand{\by}{\boldsymbol{y}}
\newcommand{\baa}{\boldsymbol{\alpha}}
\newcommand{\ti}{\textit}
\newcommand{\dd}{\delta}
\let\citep\cite 

\usepackage[caption=false]{subfig}

\usepackage{mathpazo} 
\usepackage{euler} 

\newtheorem{lemma}{Lemma}[section]

\theoremstyle{definition}

\long\def\ignore#1{}

\def\nd{\noindent}

\def\dd{\delta}

\def\bf{\textbf}
\def\ti{\textit}

\begin{document}

\title{ \Large An Efficient Scheme for Sampling in Constrained Domains}
\author{Sharang Chaudhry, Daniel Lautzenheiser, and Kaushik Ghosh}
\begin{abstract}
The creation of optimal samplers can be a challenging task, especially in the presence of constraints on the support of parameters. One way of mitigating the severity of this challenge is to work with transformed variables, where the support is more conducive to sampling. In this work, a particular transformation called inversion in a sphere is embedded within the popular Metropolis-Hastings paradigm to effectively sample in such scenarios. The method is illustrated on three domains: the standard simplex (sum-to-one constraint), a sector of an $n$-sphere, and hypercubes. The method's performance is assessed using simulation studies with comparisons to strategies from existing literature.
\end{abstract}
\address{Department of Mathematical Sciences, University of Nevada, Las Vegas, NV 89154-4020}

\Large  

\maketitle

\section{Introduction}\label{sec:intro}

\noindent The Metropolis-Hastings (MH) algorithm is a computational strategy that allows its user to sample from a target distribution when obtaining its closed-form or direct sampling is intractable. The MH algorithm is, in fact, a subclass of methods under the Markov chain Monte Carlo (MCMC) framework \cite{geyer2011introduction}. It relies on the use of a proposal distribution to generate values of a Markov chain whose stationary distribution is  the target distribution of interest. The proposed values are either accepted or rejected based on a mechanism involving the likelihood, the prior densities,  and the proposal densities. Therefore, choosing efficient proposal distributions with optimal acceptance rates is crucial for the success of any MH-sampler, which can be extremely challenging \cite{rosenthal2011adaptive}. Furthermore, the problem may become even harder when the variables of interest have constrained domains. This is because naively choosing standard proposal distributions, such as the Gaussian, may lead to poor sampling for two reasons. First, it may lead to too many rejections because proposed values are often outside the domain. Second, the proposal distribution may be too concentrated around the current value, leading to an extremely large acceptance rate. Both these issues lead to poor mixing and a potentially sub-par exploration of the parameter space.\\

\begin{figure}[h]
	\centering
	\caption{Intuition behind proposed work}
	\subfloat[][Example of a nonstandard domain $A \in \mathbb{R}^2 $ \label{fig:domA}]{\includegraphics[width=0.3\textwidth]{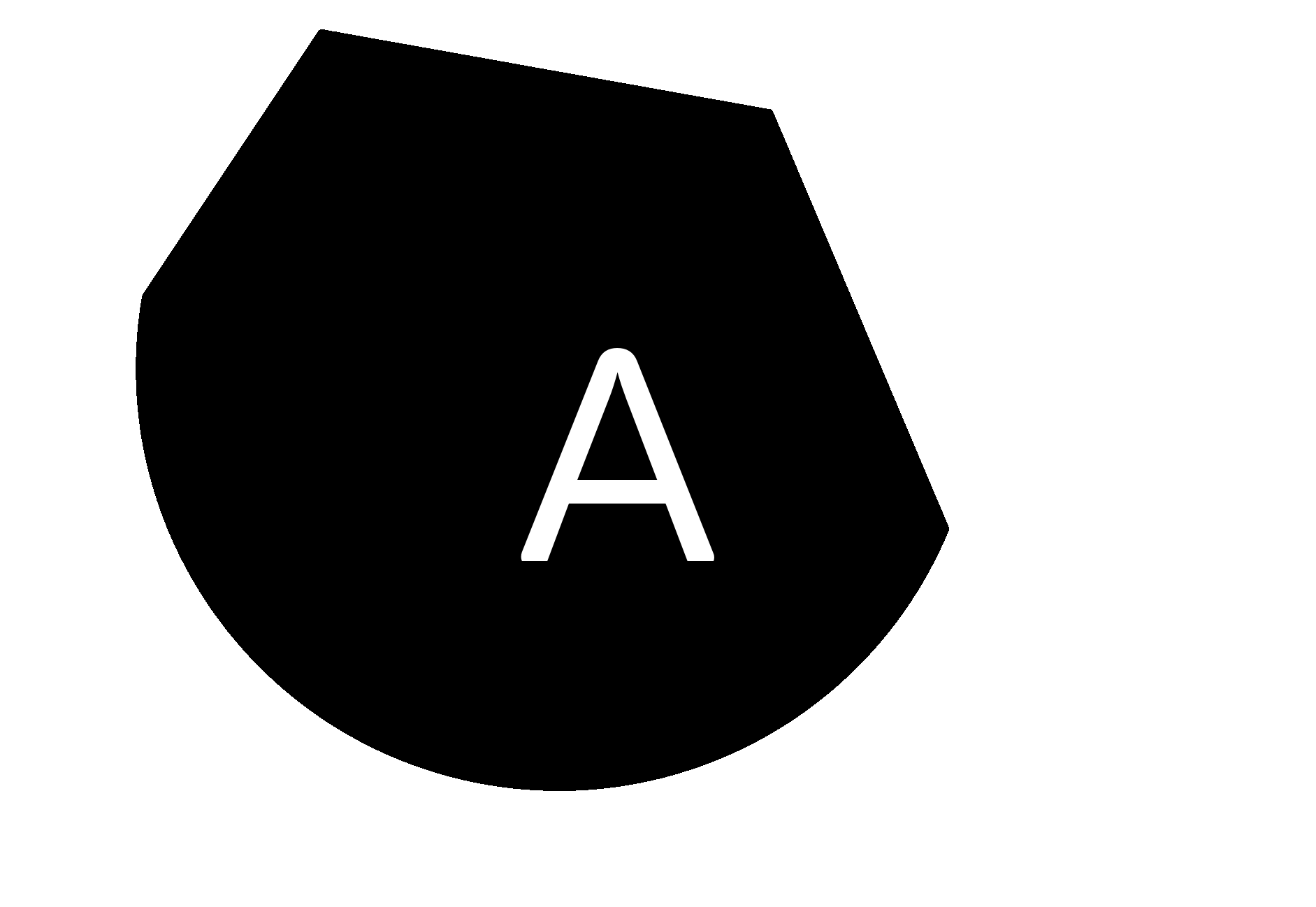}}%
	\subfloat[][Inversion in a sphere $S_{\boldsymbol{x}_0}(r)$ \label{fig:inversion}]{\includegraphics[width=0.4\textwidth]{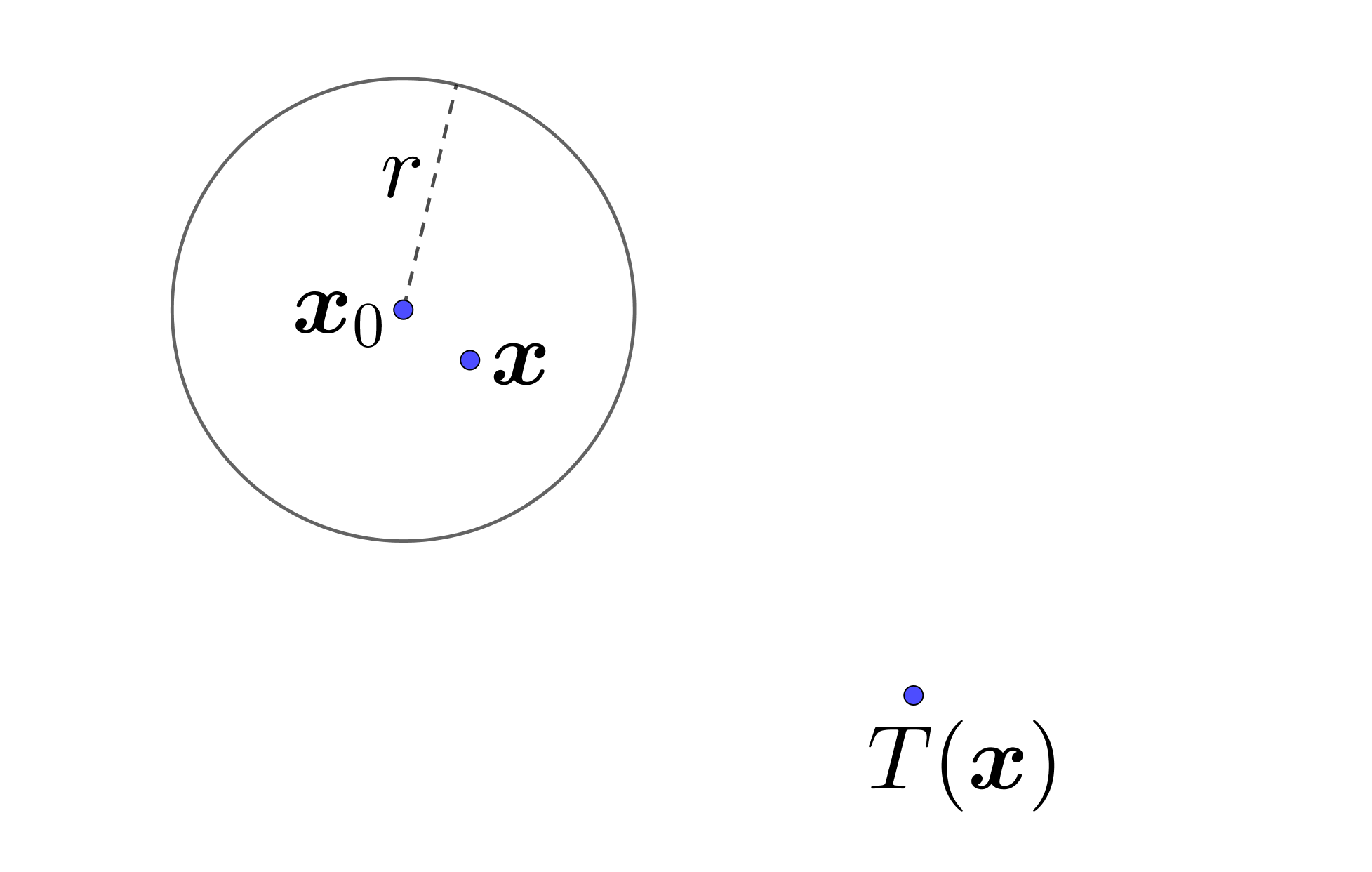}}%
	\subfloat[][Result of inverting $A$ in a sphere $S_{\boldsymbol{x}_0}(r)$ \label{fig:TA}]{\includegraphics[width=0.3\textwidth]{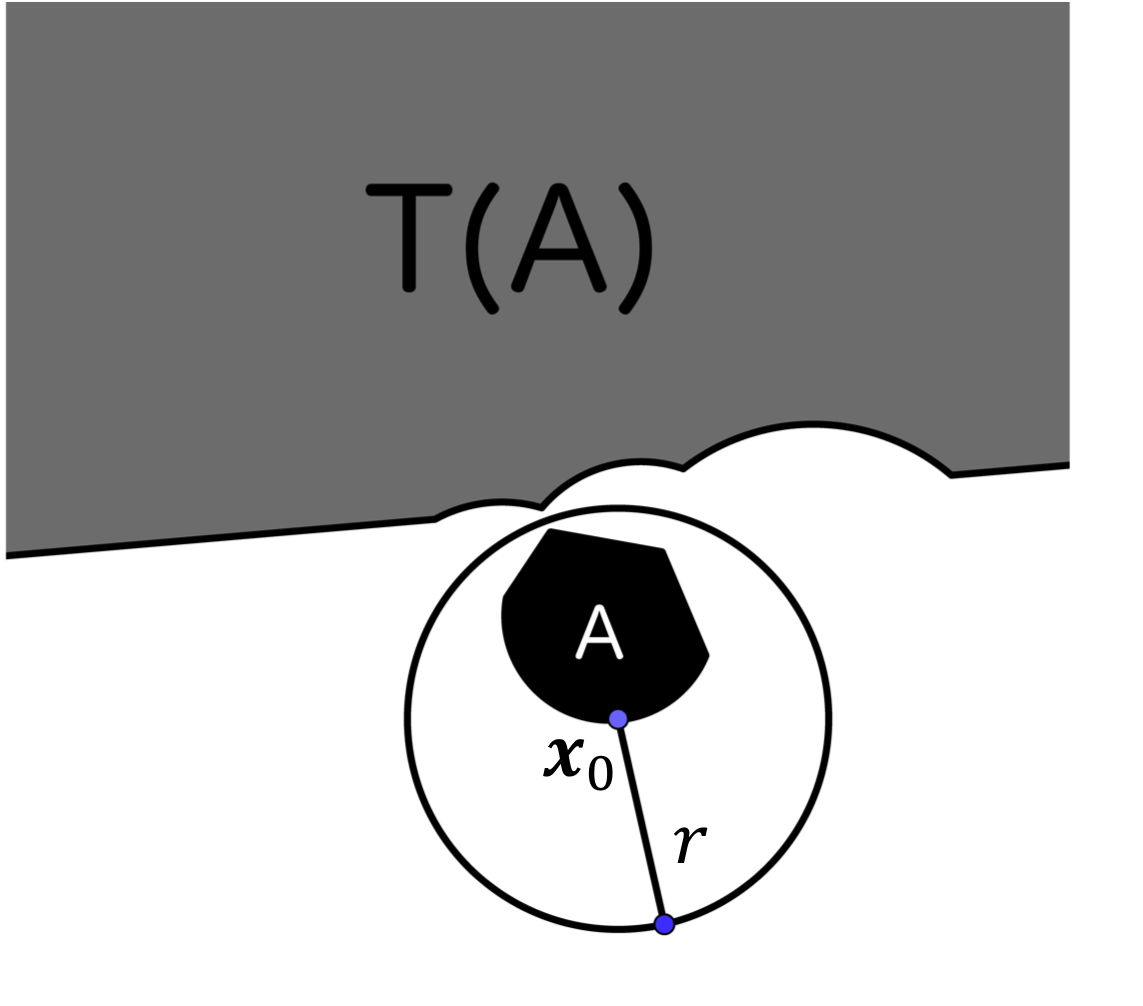}}        
	\label{fig:intuition}%
\end{figure}

\noindent  A popular strategy in situations of constrained domains involve reparametrizations such that the transformed variables have domains that are unconstrained or are of relatively large volume. The transformations that are involved in such implementations are chosen usually on a case-by-case basis (e.g., \textit{exponential} transformation for positivity), but such choices may not be entirely obvious when the domain is complicated (e.g., set $A \in \mathbb{R}^2$ as shown in Fig. \ref{fig:intuition}\subref{fig:domA}). In such situations, it is proposed to use a transformation called inversion in a sphere.\\

\noindent Given a sphere $S_{\boldsymbol{x}_0}(r) \subset \mathbb{R}^k$ with center $\boldsymbol{x}_0 \in \mathbb{R}^k$ and radius $r > 0$, inversion in $ S_{\boldsymbol{x}_0}(r) $ is a map $T: \mathbb{R}^k-\left\{ \boldsymbol{x}_0 \right\} \rightarrow \mathbb{R}^k -\left\{ \boldsymbol{x}_0 \right\} $ defined as:
\begin{equation}
\label{eq:SphInv}
T(\boldsymbol{x}) = \boldsymbol{x}_0 + r^2 \dfrac{\boldsymbol{x} - \boldsymbol{x}_0}{|| \boldsymbol{x} - \boldsymbol{x}_0 ||^2},
\end{equation}
where $ || \cdot || $ indicates the standard Euclidean norm. 
For fixed ${\boldsymbol{x}_0}$ and $r$, $T$ is an involution ($ T^{-1} = T $) that is conformal and one-to-one. The transformation is pictorially shown in Fig. \ref{fig:intuition}\subref{fig:inversion}. For its purpose in this work, $T$ maps a point inside the sphere to a point outside the sphere. Points that are closer to the center of the inverting sphere are mapped farther away from the surface of the inverting sphere. Therefore, when a region of finite volume such as the set $A$ in Fig. \ref{fig:intuition}\subref{fig:domA} is inverted in an appropriate sphere, its image $T(A)$ will occupy infinite volume as shown in Fig. \ref{fig:intuition}\subref{fig:TA}. \\

\noindent In this work, a novel scheme called Sampling Prudently using Inversion Spheres (SPInS) is presented for sampling within constrained domains. The SPInS procedure is illustrated for sampling on a simplex, in a sector of an $n$-sphere, and in a hypercube in Sections \ref{sec:simplex}, \ref{sec:nsphere}, and \ref{sec:hypercube} respectively. In each of these sections, comparative assessments are also provided. Finally, Section \ref{sec:conc} presents concluding remarks and comments on future work.
\section{Related Works}
\noindent The need for sampling in constrained distributions has garnered attention in the past few years within multiple sampling frameworks. In \cite{byrne2013geodesic} and \cite{lan2014spherical}, Hamiltonian Monte Carlo (HMC) samplers were developed for implementation on manifolds embedded within Euclidean spaces with demonstrations on unit hyperspeheres. Additionally, a derivative based importance sampler for smooth Euclidean submanifolds was presented in \cite{Chua2019}. Further, \cite{betancourt2012cruising} presented an HMC sampler specifically tailored to the simplex. \cite{altmann2014sampling} reviewed a Gibbs sampler and two HMC samplers, including the one of \cite{betancourt2012cruising}, under the assumption of Gaussianity, and \cite{smith2014gibbs} presented a detailed treatment for the mixing time of simple Gibbs samplers on the simplex. \cite{director2017efficient} presented a method to efficiently perform MH sampling on the simplex by using a Self-Adjusting Logit Transform (SALT) proposal.

\section{Sampling on the Simplex}
\label{sec:simplex}
\noindent A parameter vector $\boldsymbol{\theta}$ is said to be contained within a standard simplex in $\mathbb{R}^k$ (also known as the probability simplex) when the components of the vector have a sum-to-one constraint, that is,  
\begin{equation}
\label{eq:simplex}
\left\{ \boldsymbol{\theta} = (\theta_1,\theta_2,\dots,\theta_k) \in \mathbb{R}^k\ |\ \theta_i \geq 0, i = 1,2,\dots,k, \displaystyle \sum_{i=1}^k \theta_i= 1\right\}.
\end{equation}

\noindent The need for sampling, and eventually inference, on parameters constrained in the simplex arises in several areas including those using compositional data \cite{fry2000compositional, billheimer2001statistical, thomas2005compositional}, hyperspectral image unmixing \cite{dobigeon2009bayesian, arngren2011unmixing, bioucas2012hyperspectral, altmann2014unsupervised}, and neuroimaging \cite{behrens2007probabilistic, pisharady2018estimation} among many others.\\

\begin{figure}[h]
	\centering
	\caption{Plots of the Dirichlet densities with different scaling of the concentration parameters}\label{fig:dirichlet}
	\subfloat[][$\boldsymbol{\alpha} = ( 0.3 , 0.4  , 0.3 )$ \label{fig:alpha1}]{%
		\includegraphics[width=0.35\textwidth]{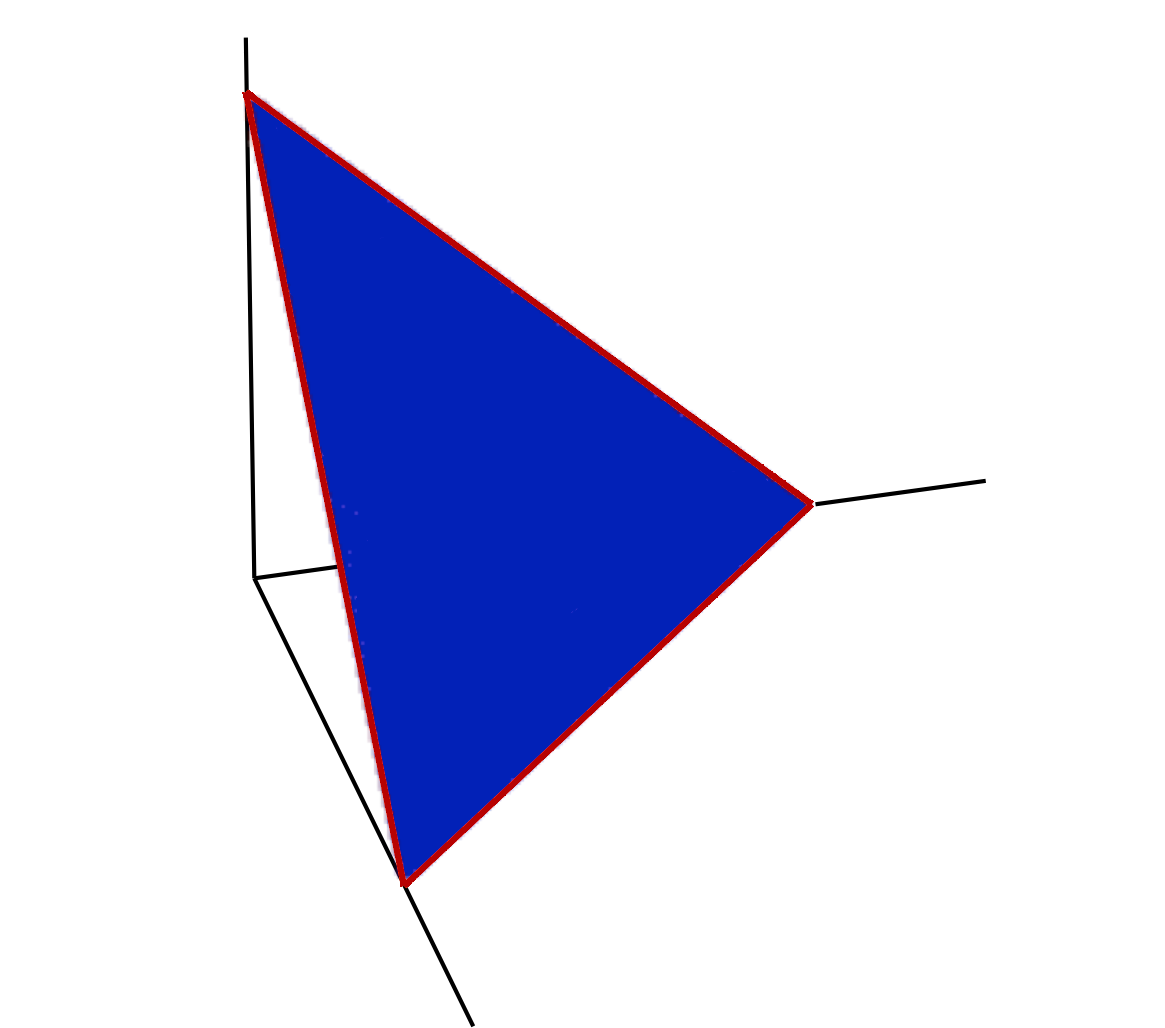}
	}        
	\subfloat[][$\boldsymbol{\alpha} = ( 3, 4 , 3)$ \label{fig:alpha10}]{%
		\includegraphics[width=0.32\textwidth]{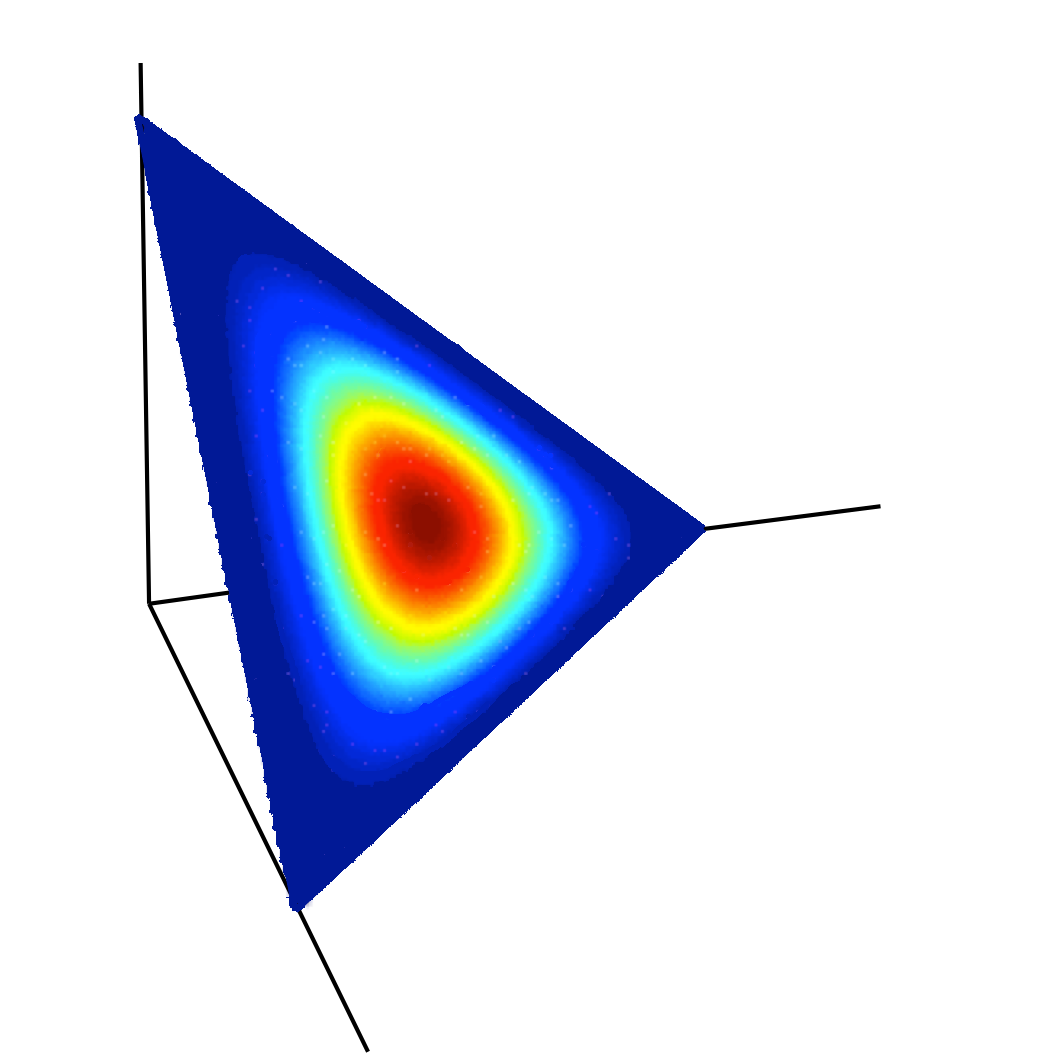}
	}        
	\subfloat[][$\boldsymbol{\alpha} = ( 30, 40 , 30)$ \label{fig:alpha100}]{%
		\includegraphics[width=0.35\textwidth]{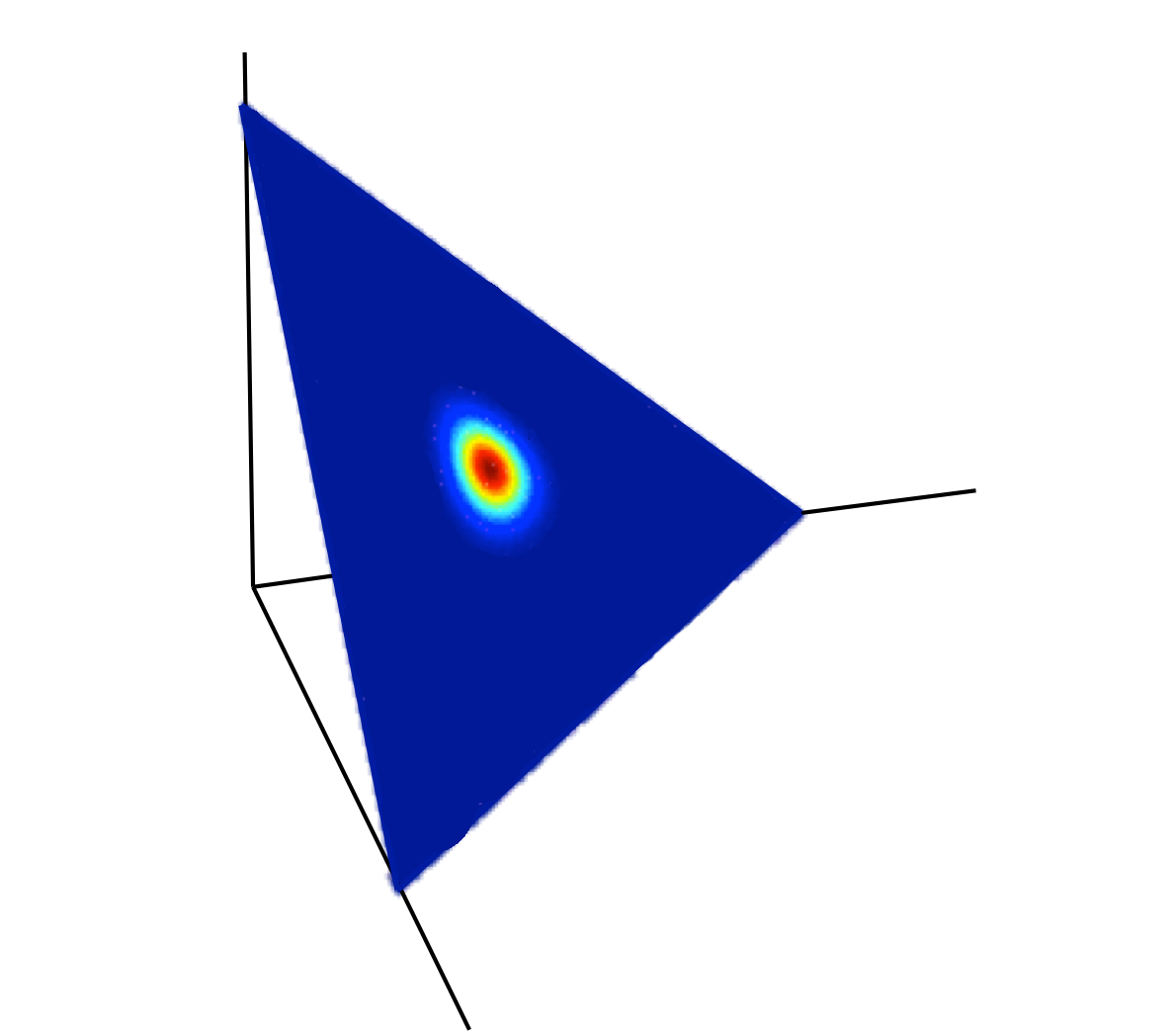}
	}        		
\end{figure}

\noindent For sampling on the simplex, the Dirichlet distribution is a natural choice for the proposal, but it cannot be used naively and requires adaptation during the course of sampling. This is because small values $(<1)$ of the shape parameters cause the density to inflate on the boundary of the simplex. As shown in Fig. \ref{fig:dirichlet}\subref{fig:alpha1}, when $\boldsymbol{\alpha} = (0.3,0.4,0.3)$, the probability of sampling in the interior of the simplex is virtually zero. Therefore, during the course of sampling, current values of the parameters (in the simplex) cannot be used as the shape parameters for the Dirichlet proposal density. However, the shape parameters may be set to a rescaled version of the current value, although it is important to understand that large values may cause the simplex to not be explored judiciously. Figs. \ref{fig:dirichlet}\subref{fig:alpha10} and \ref{fig:dirichlet}\subref{fig:alpha100} show the density as being concentrated in the interior when $\boldsymbol{\alpha} = (3,4,3)$ and $\boldsymbol{\alpha} = (30,40,30)$ respectively. The adaptation procedure employed to use the Dirichlet proposal is further discussed in Section \ref{sec:verify}.\\

\noindent  In the remainder of this section, the SPInS algorithm is presented for sampling on the simplex both in joint and componentwise manners. The performance of SPInS is then compared to SALT and an adaptive Dirichlet proposal strategy, which are direct competitors within the MH framework. Furthermore, the practicality of SPInS is demonstrated on a problem arising in neuroimaging.

\subsection{The joint proposal procedure}
\label{sec:mProposal}
\begin{figure}
	\caption{Projection of $\mathcal{S} \in \mathbb{R}^3$}
	\label{fig:3dspins}
	\hspace*{-2cm}
	\centering
	\includegraphics[scale = 0.2]{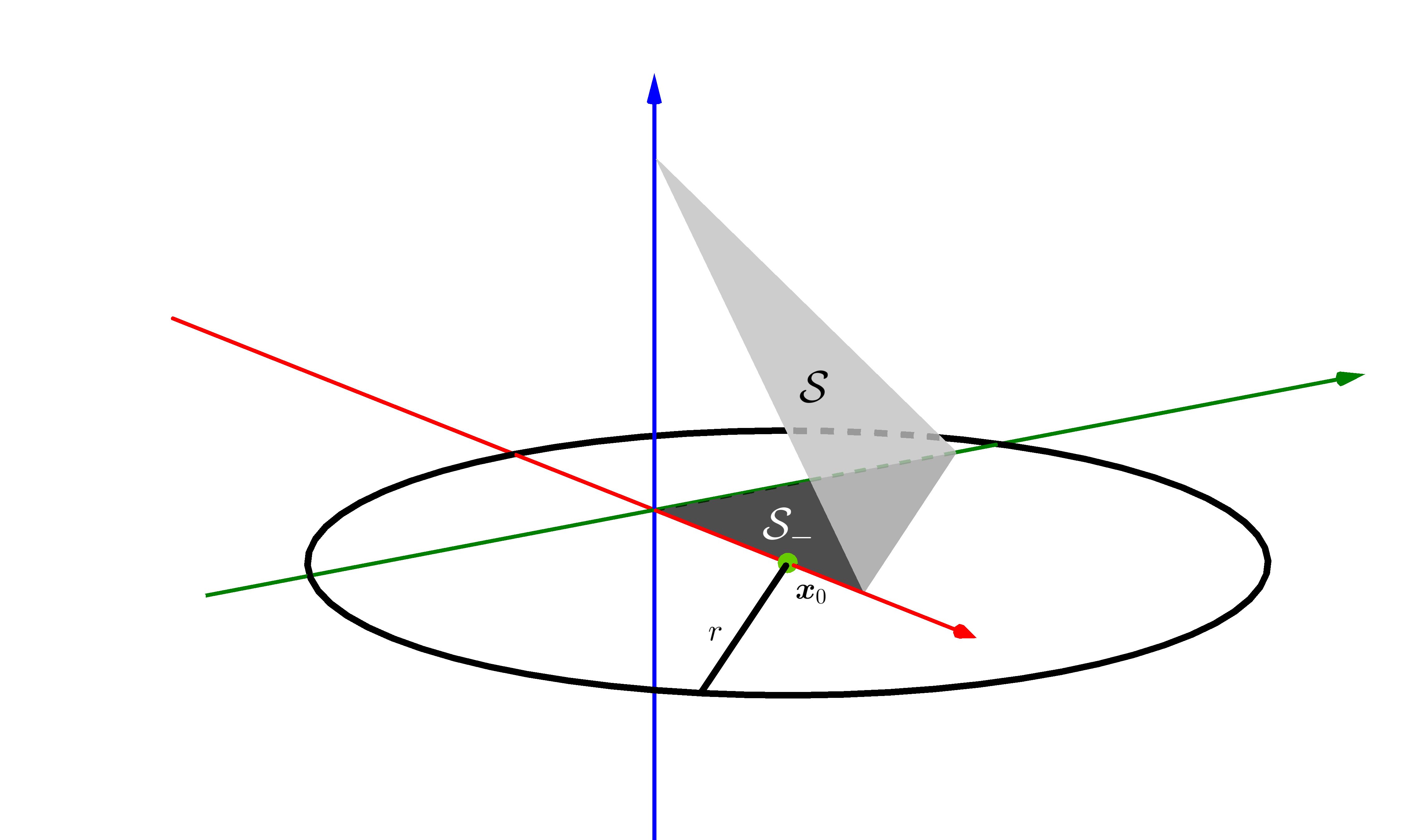}
\end{figure}

\begin{figure}[h]
	\caption{An illustration of inverting $\mathcal{S}_- \in \mathbb{R}^2$ in a sphere}
	\label{fig:3dspinsproj}
	\centering
	\includegraphics[scale = 0.45]{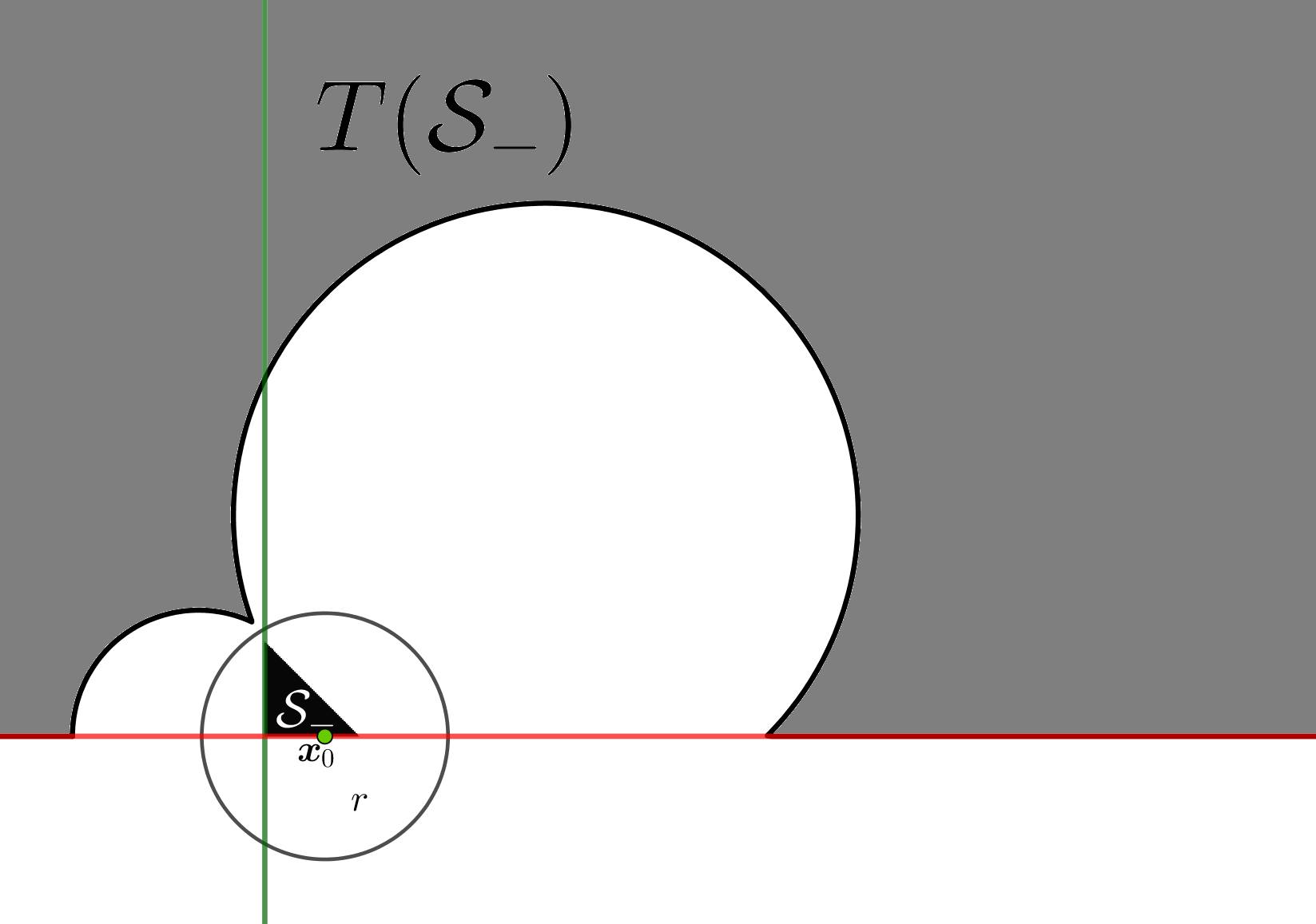}
\end{figure}

Given that $\boldsymbol{\theta} \in \mathbb{R}^k$ with a sum-to-one constraint has $k-1$ degrees of freedom, it seems intuitive to use a $k-1$ dimensional joint proposal distribution. Therefore, the simplex containing $\boldsymbol{\theta}$ is first projected down to $k-1$ dimensions by removing a single component, which can effectively be thought of as the sum to less than or equal to one constraint in $\mathbb{R}^{k-1}$. For every MH iteration, in this formulation, a random component will be removed. The following notation is first established: $\mathcal{S}$ represents the simplex of interest in $\mathbb{R}^k$, $\mathcal{S}_{-}$ is the projection of the simplex in $\mathbb{R}^{k-1}$, and $\boldsymbol{\theta}_{-} \in \mathcal{S}_{-}$ is the projection of $\boldsymbol{\theta} \in \mathcal{S}$. For example, the standard simplex in three dimensions is an equilateral triangle and projected down into two dimensions is a right triangle (See Fig \ref{fig:3dspins}). At this stage, inversion in a sphere can be performed to create an image $T(\mathcal{S}_{-})$ of infinite volume on which a standard proposal, like the Gaussian distribution, can be used (Fig \ref{fig:3dspinsproj}). \\

\noindent Although inversion in a sphere is defined naturally in a multidimensional setting, the choice of the center $\boldsymbol{x}_0$ and radius $r$ may not be obvious. It is proposed to place the center of the sphere at the closest boundary point of $T(\mathcal{S}_-)$. The advantage is that $T(\mathcal{S}_{-})$ is now of infinite volume and furthermore, points that are closer to the boundary of $\mathcal{S}_{-}$ are mapped farther away from the boundary of $T(\mathcal{S}_{-})$. This is an extremely beneficial property as it allows the proposal to adapt to the location of the current point. In essence, new proposed values remain within $T(\mathcal{S}_{-})$ and consequently, within the simplex overall. Mathematically, the center can be represented as:
\begin{equation}
\label{eqn:multcenter}
\boldsymbol{x}_0 =  \underset{\boldsymbol{x}\ \in\ \partial(\mathcal{S}_-)}{\mathrm{argmin}}\ ||\boldsymbol{\theta}_- - \boldsymbol{x} ||
\end{equation}
where $\partial(\mathcal{S}_-)$ represents the set of all boundary points of $\mathcal{S}_-$. In practice, this center is computed by finding the closest projection of $\boldsymbol{\theta}_-$ onto the faces of $\mathcal{S}_-$ as detailed in Appendix \ref{sec:app-A}. Once the center of the inversion sphere is determined, the radius of the inversion sphere can be obtained by picking a value large enough to envelope $\mathcal{S}_-$. Observing that $ \text{diam}(\mathcal{S}_-) = \sup \{ ||\boldsymbol{x} - \boldsymbol{y}||: \boldsymbol{x}, \boldsymbol{y} \in{\mathcal{S}_-}  \} = \max \{ || \boldsymbol{e}_i - \boldsymbol{e}_j||: i,j \leq k-1  \} = \sqrt{2}$, a choice of $ r = \sqrt{2}$, with $\boldsymbol{x}_0$ on the boundary of $\mathcal{S}_-$, guarantees that $ \mathcal{S}_-$ will be contained in the sphere $ S_{\boldsymbol{x}_0}\left(\sqrt{2}\right)$. With the center and radius established, the inversion in a sphere as described in (\ref{eq:SphInv}) can be performed.\\

\begin{figure}[h]
	\caption{Pictorial representation of the SPInS procedure for $\mathcal{S}_- \in \mathbb{R}^2$}
	\label{fig:3dspinspropzoom}
	\centering
	\includegraphics[scale = .35]{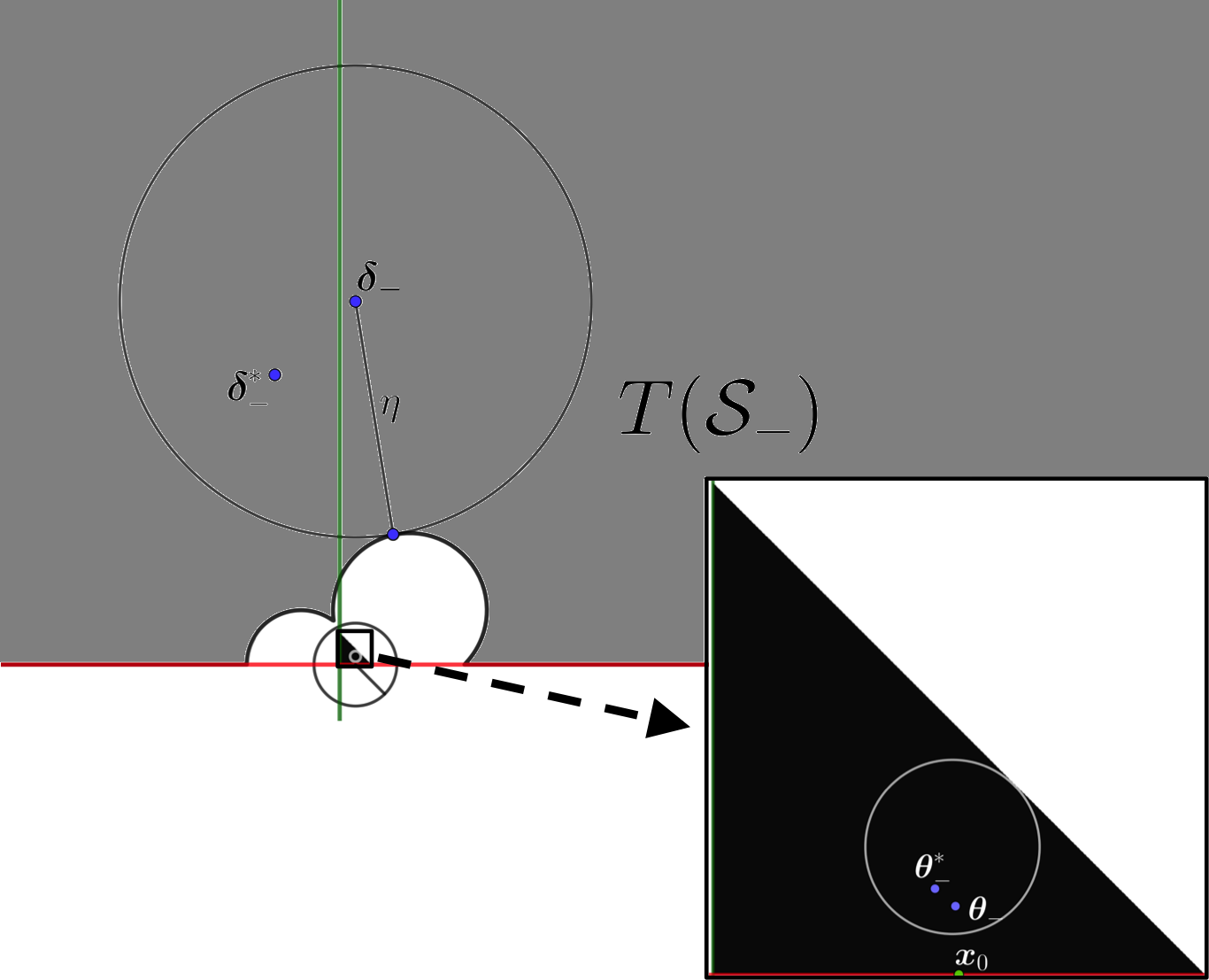}
\end{figure}

\noindent Let $T(\boldsymbol{\theta}_-) = \boldsymbol{\delta}_-$. Using $\boldsymbol{\delta}_-$ as the center, the radius ($\eta$) of the maximal ball in $ T(\mathcal{S}_-) $ is calculated (Fig. \ref{fig:3dspinspropzoom}). The purpose of computing $\eta$ is to aid in establishing a suitable covariance matrix for the proposal density that allows a new proposed value $\boldsymbol{\delta}_-^*$ to be sampled in $T(\mathcal{S}_-)$. In fact, an appropriate scaling strategy employing $\eta$ can also be used to control the mixing rate of the chain(s). An example of this is presented in Section \ref{sec:mGaussProp} when a Gaussian proposal is chosen to sample on $T(\mathcal{S}_-)$. The procedural details for finding $\eta$ are shown in Appendix \ref{sec:app-A}.   Finally, a back map $U$, a spherical inversion with center $\boldsymbol{x}_0$ and radius as chosen previously, is used to map the newly proposed value back into $\mathcal{S}_-$. Note that the realization of $U(\boldsymbol{\delta}_-^*) = \boldsymbol{\theta}_-^*$ completes the proposal procedure. A summary of the proposal procedure is shown in (\ref{eq:spinsproc}) and pictorially presented in Fig. \ref{fig:3dspinspropzoom}.

\begin{equation}
\label{eq:spinsproc}
\boldsymbol{\theta} \longrightarrow \boldsymbol{\theta}_- \longrightarrow T(\boldsymbol{\theta}_-) = \boldsymbol{\delta}_- \xrightarrow{q(\cdot|\boldsymbol{\delta}_-)} \boldsymbol{\delta}_-^* \longrightarrow U(\boldsymbol{\delta}_-^*) = \boldsymbol{\theta}_-^* \longrightarrow \boldsymbol{\theta}^*.
\end{equation}

\subsubsection{Joint SPInS-proposal density}
\label{sec:mGaussProp}
It is now important to derive the proposal density $q(\boldsymbol{\theta}^*|\boldsymbol{\theta})$ for the choice of $q(\boldsymbol{\delta}^*_-|\boldsymbol{\delta}_-)$. It should be noted that the transformations $\boldsymbol{\theta} \longrightarrow  \boldsymbol{\theta}_-$ and $\boldsymbol{\theta}_-^* \longrightarrow \boldsymbol{\theta}^*$ are deterministic. Specifically, the first involves removal of a known component and the second involves the missing component being computed by subtracting the sum of the other components from one. This leads to the fact that $q(\boldsymbol{\theta}^*|\boldsymbol{\theta}) = q(\boldsymbol{\theta}^*_-|\boldsymbol{\theta}_-)$.\\ 

\noindent Next, note that the choice of $\boldsymbol{x}_{0}$ depends on $\boldsymbol{\theta}$ implying that the choice of the inversion sphere changes with the current location. Therefore, the properties of a standard spherical inversion transformation as presented in Section \ref{sec:intro}, where the inversion sphere is established a priori, may need to be reexamined. The key property required for the computation of the proposal density is that $T$ is one-to-one. This is proved in Appendix \ref{sec:app-B}, and leads to the fact that $q(\boldsymbol{\delta}^*_-|\boldsymbol{\delta}_-) = q \left (\boldsymbol{\delta}^*_-|\boldsymbol{T(\theta}_-)\right)$, which further implies $q(\boldsymbol{\delta}^*_-|\boldsymbol{\theta}_-)$ is of known form. Finally, $q(\boldsymbol{\theta}^*_-|\boldsymbol{\theta}_-) = q\left(U(\boldsymbol{\delta}^*)_-|\boldsymbol{\theta}_-\right)$ can be easily obtained by the using the Jacobian transformation method for random variables since $U$ is deterministic and is a standard inversion in a sphere. More importantly, note that $U$ is in fact $T^{-1}$ when the domain of $T$ is restricted to $\mathcal{S}_-$.

\subsubsection*{Proposal density using a Gaussian distribution}
\label{sec:mGaussProp}
The Gaussian proposal density on $T(\mathcal{S}_-)$ can be represented as:
\begin{equation}
\begin{split}
q(\boldsymbol{\delta}_-^*|\boldsymbol{\delta}_-) &= \dfrac{1}{\sqrt{(2 \pi)^{k-1} |\Sigma| }} \exp\left\{ -\dfrac{1}{2}\left(\boldsymbol{\delta}_-^*-\boldsymbol{\delta}_-\right)^t\Sigma^{-1} \left(\boldsymbol{\delta}_-^*-\boldsymbol{\delta}_-\right)  \right\}. \\ 
\implies q(\boldsymbol{\delta}_-^*|\boldsymbol{\theta}_-) &= \dfrac{1}{\sqrt{(2 \pi)^{k-1} |\Sigma| }} \exp\left\{ -\dfrac{1}{2}(\boldsymbol{\delta}_-^*-T(\boldsymbol{\theta}_-))^t\Sigma^{-1} (\boldsymbol{\delta}_-^*-T(\boldsymbol{\theta}_-)  )\right\}.
\end{split}
\end{equation}

\noindent In the above, $\Sigma = \left(\dfrac{\eta}{d} \right)^ 2 I_{k-1}$, where $I_{k-1}$ is the $(k-1) \times (k-1)$ identity matrix, and the parameter $d$ is chosen to appropriately scale the Gaussian proposal to both propose values within the simplex and adjust acceptance rates as necessary. The overall proposal density can be written as:
\begin{equation}
\small
\begin{split}
q(\boldsymbol{\theta}_-^*|\boldsymbol{\theta}_-)  = & \dfrac{1}{\sqrt{(2 \pi)^{k-1} |\Sigma| }} \exp\left\{ -\dfrac{1}{2}\left(U^{-1}(\boldsymbol{\theta}_-^*)-T(\boldsymbol{\theta}_-)\right)^t\Sigma^{-1} \left(U^{-1}(\boldsymbol{\theta}_-^*)-T(\boldsymbol{\theta}_-)  \right)\right\}  |J| \\
= & \dfrac{1}{\sqrt{(2 \pi)^{k-1} |\Sigma| }}  \exp\left\{ -\dfrac{1}{2}\left(U(\boldsymbol{\theta}_-^*)-T(\boldsymbol{\theta}_-)\right)^t\Sigma^{-1} \left(U(\boldsymbol{\theta}_-^*)-T(\boldsymbol{\theta}_-)  \right)\right\} \left(\dfrac{r}{\displaystyle ||\boldsymbol{\theta}_-^* - \boldsymbol{x}_0|| }\right)^{2(k-1)} \\
= & q(\boldsymbol{\theta}^*|\boldsymbol{\theta}).
\end{split}
\end{equation}
The derivation of the Jacobian is given in Appendix \ref{sec:app-B}. It should be noted that the second line of the equality is obtained from the fact that $U$ is an involution. Finally, any values proposed outside $T(\mathcal{S}_-)$ are automatically rejected.

\subsection{The componentwise proposal procedure}
\label{sec:Proposal}
In addition to the proposal procedure shown in the previous section, a componentwise MH update strategy can also be developed. A single iteration, here, will consist of $k$-individual MH steps, each corresponding to a different component in $\boldsymbol{\theta}$. For each component $\theta_i\ \text{of}\  \boldsymbol{\theta}\  (i = 1,2,\dots,k)$, the inversion sphere described in Section \ref{sec:intro} will be an interval.\\

\noindent For a given component $\theta_i \in [0,1]$, a sphere $S_{x_{0,i}}(r)$, is established with the same principles as the joint sampler. The closest endpoint of the unit interval is chosen as the center, $ x_{0,i} $ and a radius of $ r = 1 $ are selected because this guarantees that the entire interval, (0,1),  is encapsulated within the inversion sphere. More specifically, 
\begin{equation}
x_{0,i} = \begin{cases} 0 & \text{if}\ \theta_i \leq \frac{1}{2}, \\ 1 & \text{if}\ \theta_i > \frac{1}{2}, \end{cases}\quad S_{x_{0,i}}(r) = \begin{cases} (-1 , 1) & \text{if}\ \theta_i \leq \frac{1}{2}, \\ (0,2) & \text{if}\ \theta_i > \frac{1}{2}. \end{cases}
\end{equation}
The inversion in this sphere can then be performed by using the map $T$ as below.
\begin{equation}
\label{Tmap}
T(\theta_i) = \begin{cases} \frac{1}{\theta_i} & \text{if}\ \theta_i \leq \frac{1}{2}, \\ \frac{\theta_i}{\theta_i-1} & \text{if}\ \theta_i > \frac{1}{2}. \end{cases}
\end{equation} 

\noindent It is easy to see that T is one-to-one. Furthermore, if $\theta_i \rightarrow 0^+$, then $ T(\theta_i) \rightarrow \infty $ and if $ \theta_i \rightarrow 1^- $, then $ T(\theta_i) \rightarrow - \infty $. This means that neighborhoods of points in the $ T $-image of the unit interval may now occupy a considerably larger length in comparison to the unit interval itself and follows the same rationale outlined in Sections \ref{sec:intro} and \ref{sec:mProposal}.\\

\noindent Let  $T(\theta_i) = \delta_i$. Around $\delta_i$, the maximal interval that fits within the image of the unit interval, $T\left((0,1)\right)$, is calculated. Let the radius of this interval be called $\eta_i$, given by:
\begin{equation}
\begin{split}
\eta_i & = \begin{cases} T(\theta_i)-1 & \text{if}\ \theta_i \leq \frac{1}{2}, \\ |T(\theta_i)| & \text{if}\ \theta_i > \frac{1}{2} \end{cases}  \\ 
& = \begin{cases} \frac{1}{\theta_i}-1 & \text{if}\ \theta_i \leq \frac{1}{2}, \\ |\frac{\theta_i}{\theta_i-1}| & \text{if}\ \theta_i > \frac{1}{2} \end{cases} \\
& = \begin{cases} \frac{1-\theta_i}{\theta_i} & \text{if}\ \theta_i \leq \frac{1}{2}, \\ \frac{\theta_i}{1-\theta_i} & \text{if}\ \theta_i > \frac{1}{2}. \end{cases}
\end{split}
\end{equation}
As in the joint formulation with $\eta$, the purpose of $\eta_i$ is to help appropriately define the support for the proposal distribution on $T\left((0,1)\right)$ and aid in controlling the mixing proportions of the individual chains. On $T\left((0,1)\right)$, a proposal density, $q(\cdot|\delta_i)$, of choice can be used to sample a new point $\delta_i^*$. With a new sampled point, a back mapping, denoted by $U$, is used to obtain the point's corresponding pre-image on the unit interval. This map corresponds to an inversion in the sphere $S_{x_{0,i}}(r)$ as chosen at the initial step. It should be noted again that $U$ is in fact $T^{-1}$ when the domain of T is restricted to $(0,1)$.
\begin{equation}
\label{U-map}
U(\delta^*_i)  = \begin{cases}  \frac{1}{\delta^*_i} & \text{if}\ \theta_i \leq \frac{1}{2}, \\ \frac{\delta^*_i}{\delta^*_i-1} & \text{if}\ \theta_i > \frac{1}{2}. \end{cases}
\end{equation}

\begin{figure}[h]
	\centering
	\caption{Overview of proposal procedure}
	\label{fig:ComponentSampling}
	\includegraphics[scale = 0.5]{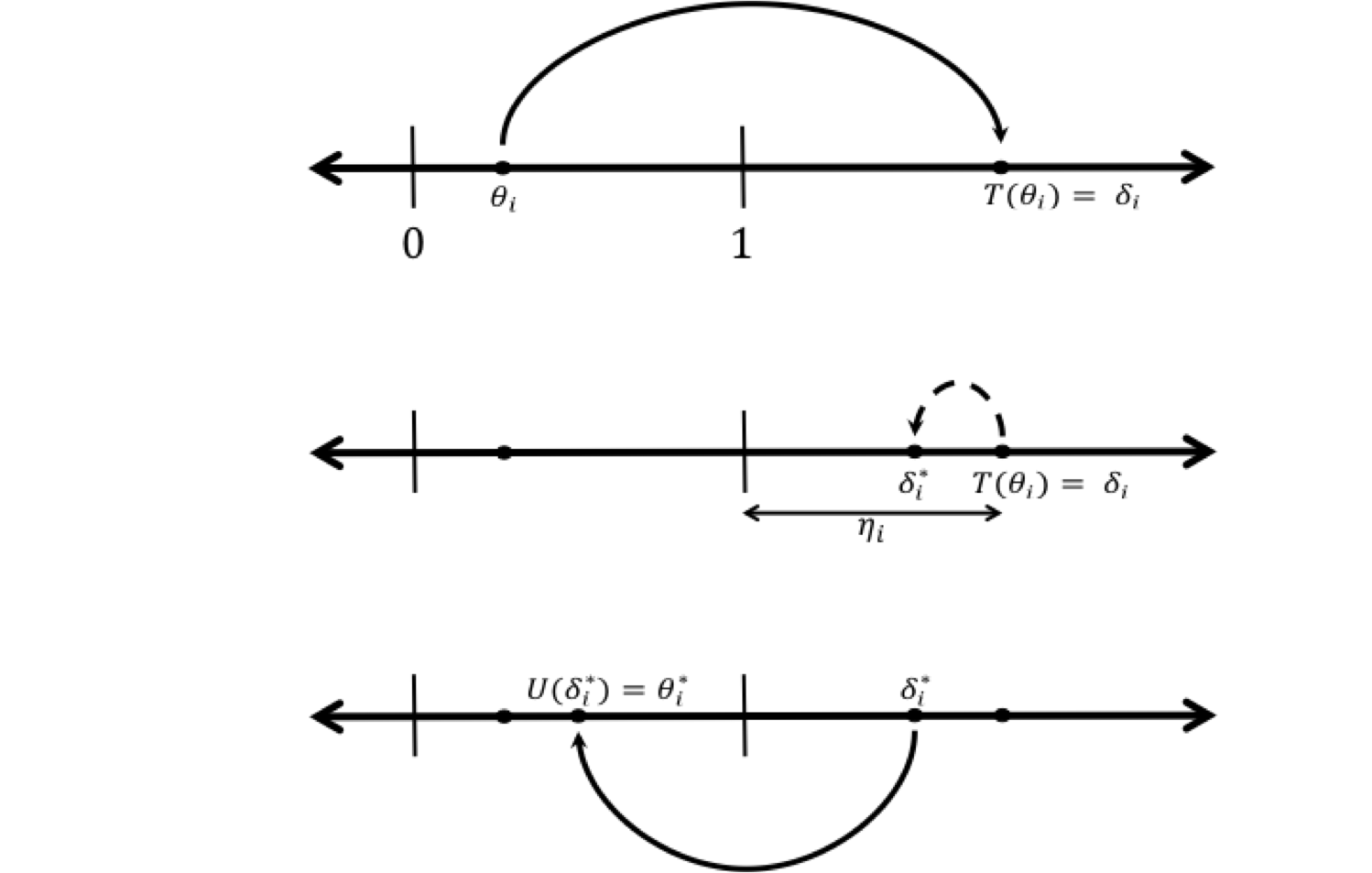}
\end{figure}

\noindent The overall procedure used to obtain a new proposal point is exemplified pictorially in Fig. \ref{fig:ComponentSampling} and can be written as follows:
\begin{equation}
\theta_i \longrightarrow T(\theta_i) = \delta_i \xrightarrow{q(\cdot|\delta_i)} \delta_i^* \longrightarrow U(\delta_i^*) = \theta_i^* .
\end{equation}

\noindent Note that $\eta_i \rightarrow \infty$  as $\theta_i \rightarrow 0^+$ or $\theta_i \rightarrow 1^-$. It should be reiterated that this property is desirable because it allows the proposals to dynamically adjust based on the location of the point $\theta_i$ in the unit interval. Ordinarily, no such adjustment would have taken place, and for points close to 0 or 1 sampling may become inefficient as discussed in Section \ref{sec:intro}. The componentwise SPInS-Gaussian proposal density is derived next in Section \ref{sec:GaussProp}

\subsubsection{Componentwise SPInS-proposal density}
\label{sec:SPInS-proposaldensity}
The proposal density $q(\theta^*_i|\theta_i)$ is obtained using ideas similar to the previous section. Since $T$ is a one-to-one mapping, the choice $q(\delta^*_i|\delta_i) = q(\delta^*_i|T(\theta_i))$ implies that $q(\delta^*_i|\theta_i)$ is of a known form. Furthermore, $q(\theta^*_i|\theta_i) = q(U(\delta^*_i)|\theta_i)$ can be obtained by simply using the Jacobian transformation method for random variables. It is critical to mention that a change in any one component necessitates a readjustment in other components due to the sum-to-one constraint. In order to do so, each component of $\left\{\theta_j\right\}_{j \neq i}$ is rescaled in a manner which maintains relative ratios.

\begin{equation}
\theta_j^* = (1-\theta_i^*)\ \dfrac{\theta_j}{1-\theta_i}, j \neq i.
\end{equation}

\noindent This rescaling is deterministic and therefore, doesn't play a role in the computation of the proposal density.

\subsubsection*{Proposal density using a Gaussian distribution}
\label{sec:GaussProp}
The choice of using a Gaussian proposal on the image of the unit interval means that
\begin{equation}
\begin{split}
q(\delta^*_i|\delta_i) &= \dfrac{1}{\sqrt{2 \pi (\frac{\eta_i}{d})^2}}\ \exp\left\{ \dfrac{-(\delta^*_i - \delta_i)^2}{2 (\frac{\eta_i}{d})^2 }\right\}. \\ 
\implies q(\delta^*_i|\theta_i) &= \dfrac{1}{\sqrt{2 \pi (\frac{\eta_i}{d})^2}}\ \exp\left\{ \dfrac{-(\delta^*_i - T(\theta_i))^2}{2 (\frac{\eta_i}{d})^2 }\right\}.
\end{split}
\end{equation}

\noindent The parameter $d$ above is interpreted in exactly the same manner as the joint sampler explored previously. Next, using the fact that $U$ is an involution, the SPInS-Gaussian density can be written as follows.
\begin{equation}
\label{propdens}
\begin{split}
q(\theta^*_i|\theta_i) &= \dfrac{1}{\sqrt{2 \pi (\frac{\eta_i}{d})^2}}\ \exp\left\{ \dfrac{-(U^{-1}(\theta_i^*) - T(\theta_i))^2}{2 (\frac{\eta_i}{d})^2 }\right\} \left| \dfrac{d\ U^{-1}(\theta_i^*)}{d\ \theta_i^*}\right| \\
& = \dfrac{1}{\sqrt{2 \pi (\frac{\eta_i}{d})^2}}\ \exp\left\{ \dfrac{-(U(\theta_i^*) - T(\theta_i))^2}{2 (\frac{\eta_i}{d})^2 }\right\} \left|\dfrac{d\ U(\theta_i^*)}{d\ \theta_i^*}\right|. 
\end{split}
\end{equation}

\noindent  The Jacobian term can then be obtained easily from (\ref{U-map}).
\begin{equation}
\left|\dfrac{d\ U(\theta_i^*)}{d\ \theta_i^*}\right| = \begin{cases} \ \left|\frac{-1}{(\theta^*_i)^2}\right| & \text{if}\ \theta_i \leq \frac{1}{2}, \\ \left| \frac{-1}{(\theta^*_i-1)^2} \right| & \text{if}\ \theta_i > \frac{1}{2} \end{cases} = \begin{cases}  \frac{1}{(\theta^*_i)^2} & \text{if}\ \theta_i \leq \frac{1}{2}, \\ \frac{1}{(\theta^*_i-1)^2} & \text{if}\ \theta_i > \frac{1}{2}. \end{cases}
\end{equation}

\noindent Note when $d=3$, there is only a 0.15\% chance that any proposed value will fall outside the image of the simplex. These values, if proposed, will be automatically rejected.

\subsection{Simulation Studies}
\label{sec:verify}
In this section, datasets are simulated from two hypothetical generative models for demonstration and testing. For comparative analysis, both formulations of the SPInS (joint and componentwise) along with SALT and the adaptive Dirichlet MH sampler are utilized for posterior sampling. The prior distribution in both these examples and for all samplers is assumed to be $Dirichlet(1,1,1)$. The ``betterness" of the samplers in the context of this work is quantified in terms of mixing time. In other words, how quickly do the Markov Chains explore the domain and converge.\\

\noindent As outlined initially in Section \ref{sec:simplex}, the Dirichlet proposal requires adaptation to be effective for sampling. This adaptation could come in the form of a rescaling of the current value, i.e., $\boldsymbol{\theta}^* \sim Dirichlet(\lambda \boldsymbol{\theta} )$. To ensure that all rescaled values are at least one, $\lambda$ must be chosen to be greater than the multiplicative inverse of the smallest component of $\boldsymbol{\theta}$. In this work, $\lambda$ is determined by first choosing its appropriate order of magnitude and then finding another scalar to achieve optimal mixing rates. For example, if the current value of $\boldsymbol{\theta} = (0.01, 0.001, 0.1, 0.889)$ then $\lambda = \tau \times 10^3$. Note that $\tau$ is heuristically determined and can be thought of as an analog to the scale parameter when using Gaussian proposals. \\

\subsubsection{Additive Correlated Error}
\label{sec:msnerror}
\noindent Let $\boldsymbol{\theta} \in \mathbb{R}^3$ be some unknown fixed value and noisy signals $\mathbf{y} = \boldsymbol{\theta} + \boldsymbol{\epsilon}$ be observed. Here, $\boldsymbol{\epsilon} \sim MSN(\boldsymbol{\xi},\Omega,\boldsymbol{\alpha})$ and $MSN$ refers to a Multivariate Skewed Normal distribution characterized by its location $(\boldsymbol{\xi})$, scale matrix $(\Omega)$, and slant $(\boldsymbol{\alpha})$. More information on the $MSN$ distribution can be found in \cite{azzalini2013skew}. The $MSN$ noise is added instead of the more conventional Gaussian noise for two reasons; a) to show the flexibility of MH sampling and b) under the assumption of Gaussian noise, Gibbs sampling is possible since the conditional posterior turns out to be a truncated multivariate Gaussian distribution.\\

\noindent A dataset containing 1000 observations is simulated by setting $\boldsymbol{\theta} = \left(\frac{1}{3},\frac{1}{3},\frac{1}{3}\right) $ with $\boldsymbol{\xi} = \boldsymbol{0}$, $\Omega = \begin{bmatrix} 6&-3 &3 \\-3 &3 &0 \\ 3&0 & 6\end{bmatrix}$, and $\boldsymbol{\alpha} = (1,1,1)$. For the evaluation of the likelihood function, the parameters of the $MSN$ distribution are considered to be known. All four samplers were run multiple times to determine appropriate scaling parameters that guarantees good mixing (acceptance rate $\sim 40\%$). For the componentwise and joint SPInS samplers, $d$ was found to be 2.5 and 3 respectively. For the SALT sampler, $h$ representing an inverse analog of $d$ was determined to be $0.4$. Finally, for the adaptive Dirichlet sampler, $\tau$ was set to $10$. A run of each sampler initialized at $(10^{-10},10^{-10}, 9.999999998\times10^{-1})$, which given the geometry of the simplex and the true value of $\boldsymbol{\theta}$ is an extremely poor starting value, is presented below. The purpose of choosing this starting value was to examine how many iterations the samplers take to converge.\\

\begin{figure}[h]
	\centering
	\caption{Posterior sample for additive $MSN$ noise}\label{fig:pos-msn}
	\subfloat[][SPInS componentwise sampler \label{fig:msn-spins-comp}]{%
		\includegraphics[width=0.5\textwidth]{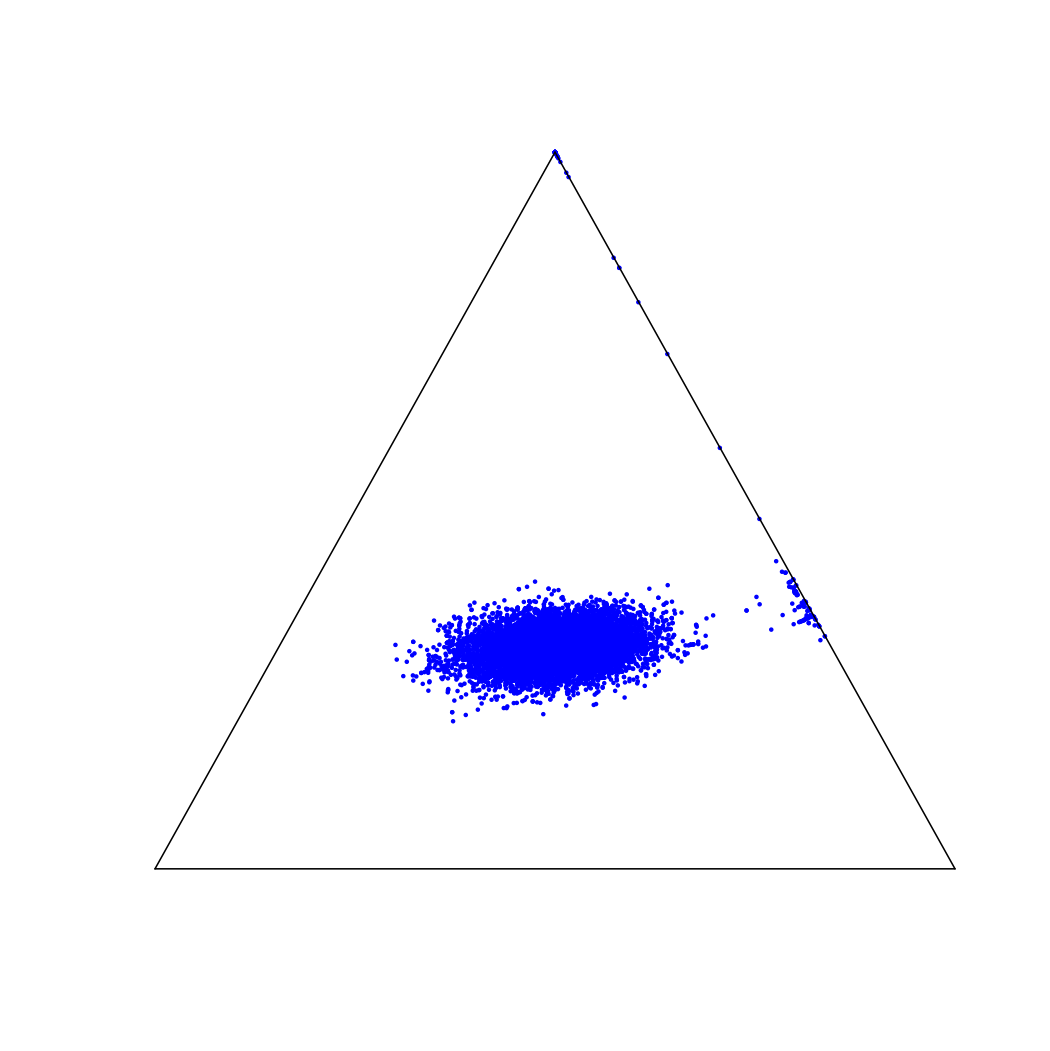}
	} 
	\subfloat[][SPInS joint sampler \label{fig:msn-spins-mult}]{%
		\includegraphics[width=0.5\textwidth]{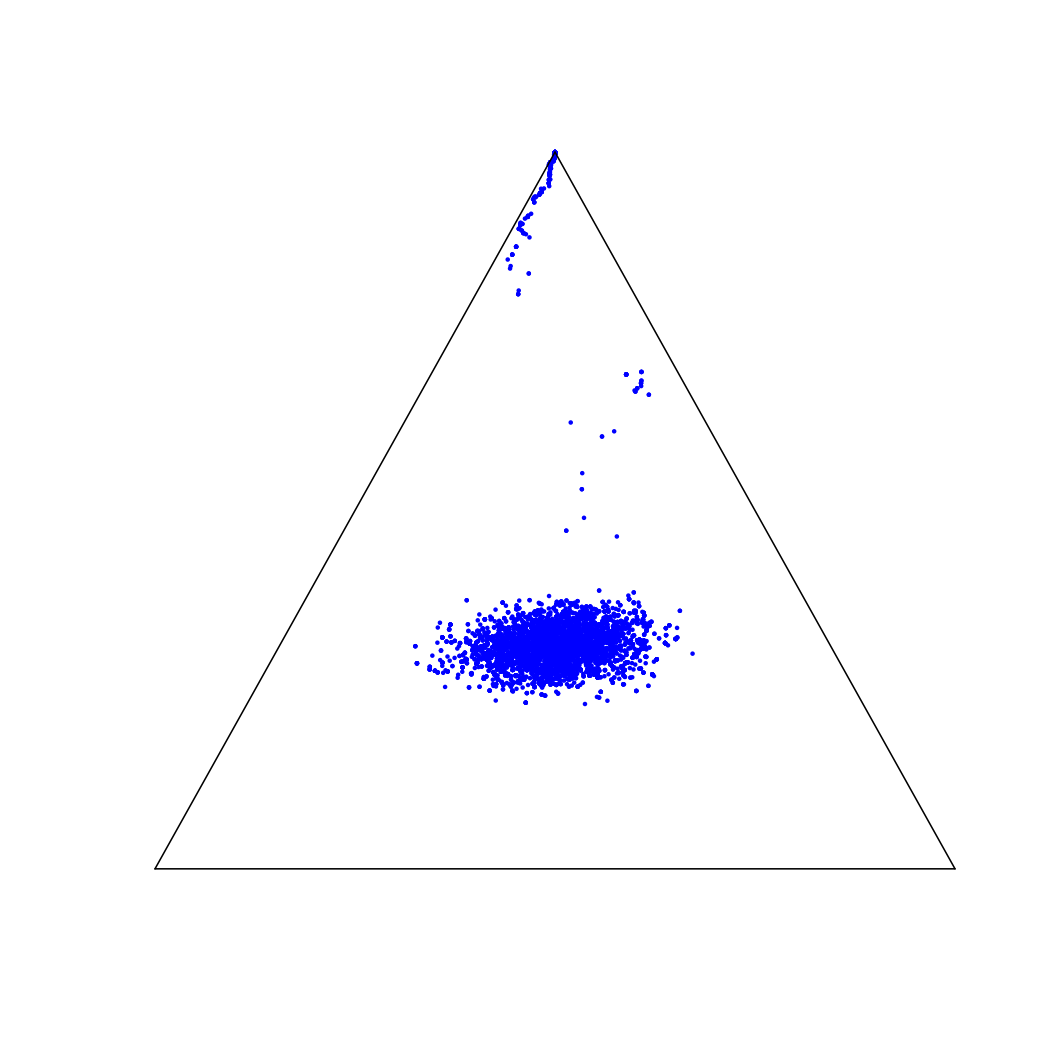}
	} \\
	\subfloat[][SALT sampler \label{fig:msn-salt}]{%
		\includegraphics[width=0.5\textwidth]{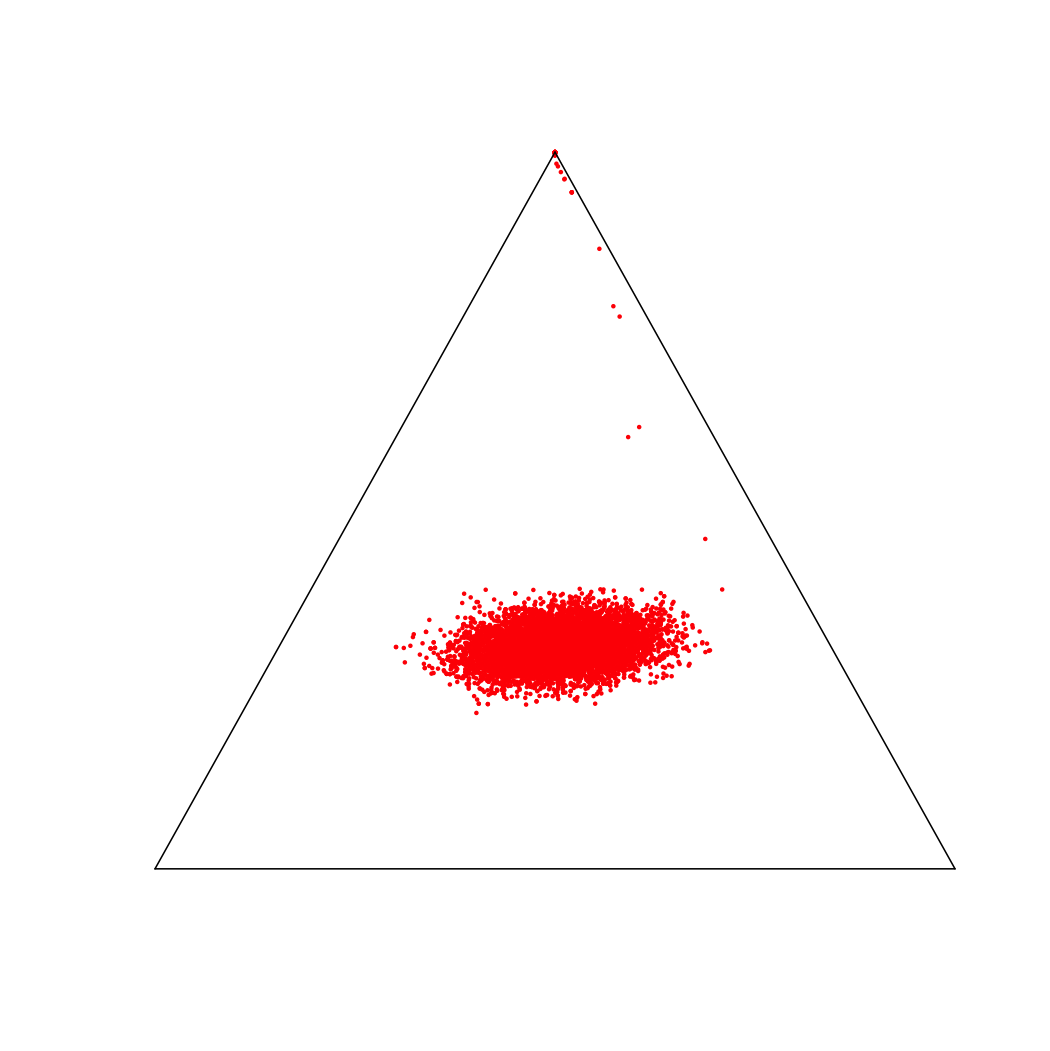}
	} 
	\subfloat[][Adaptive Dirichlet sampler \label{fig:msn-dirichlet}]{%
		\includegraphics[width=0.5\textwidth]{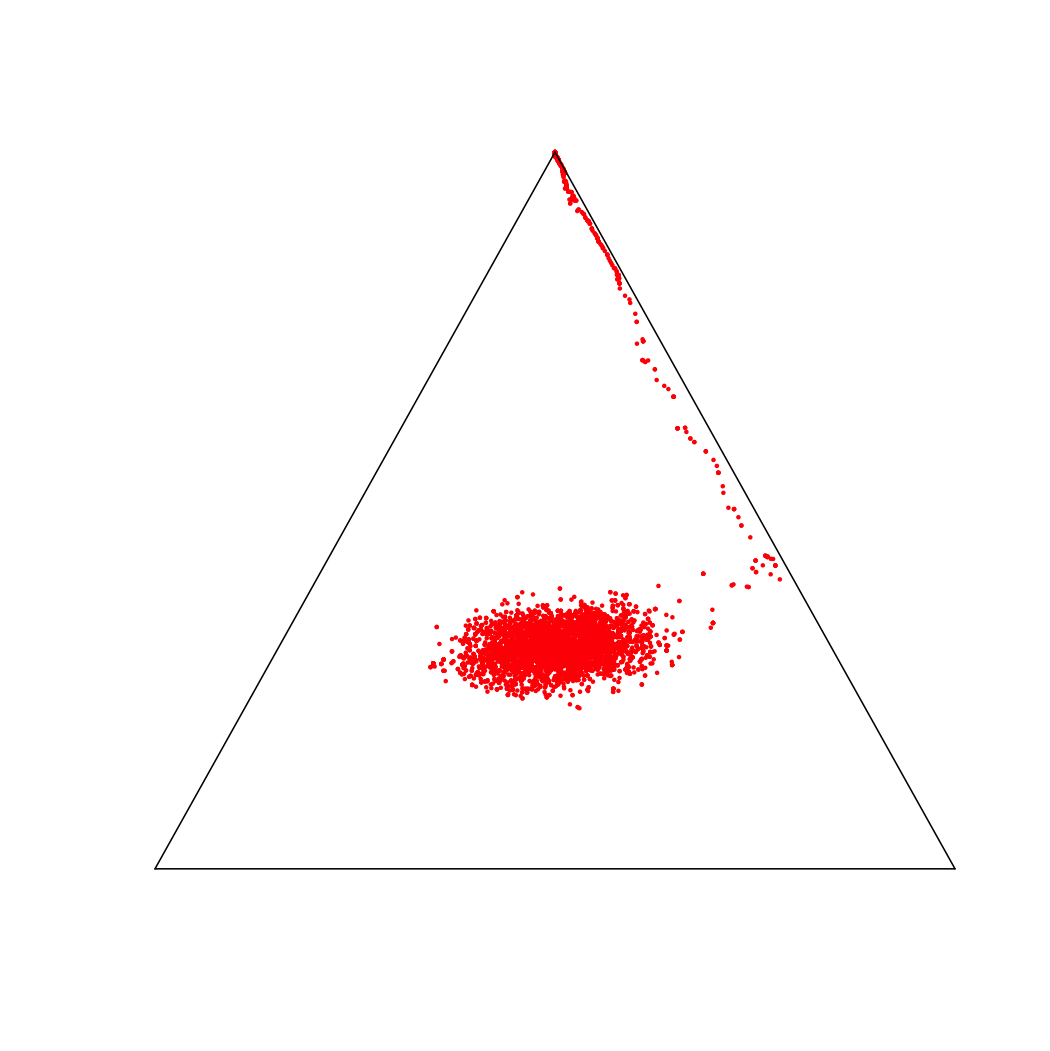}
	}
\end{figure}

\begin{figure}[h]
	\centering
	\caption{Trace plots for additive $MSN$ noise}\label{fig:pos-msn-tr}
	\subfloat[][SPInS componentwise sampler \label{fig:msn-spins-comp-tr}]{%
		\includegraphics[width=0.5\textwidth]{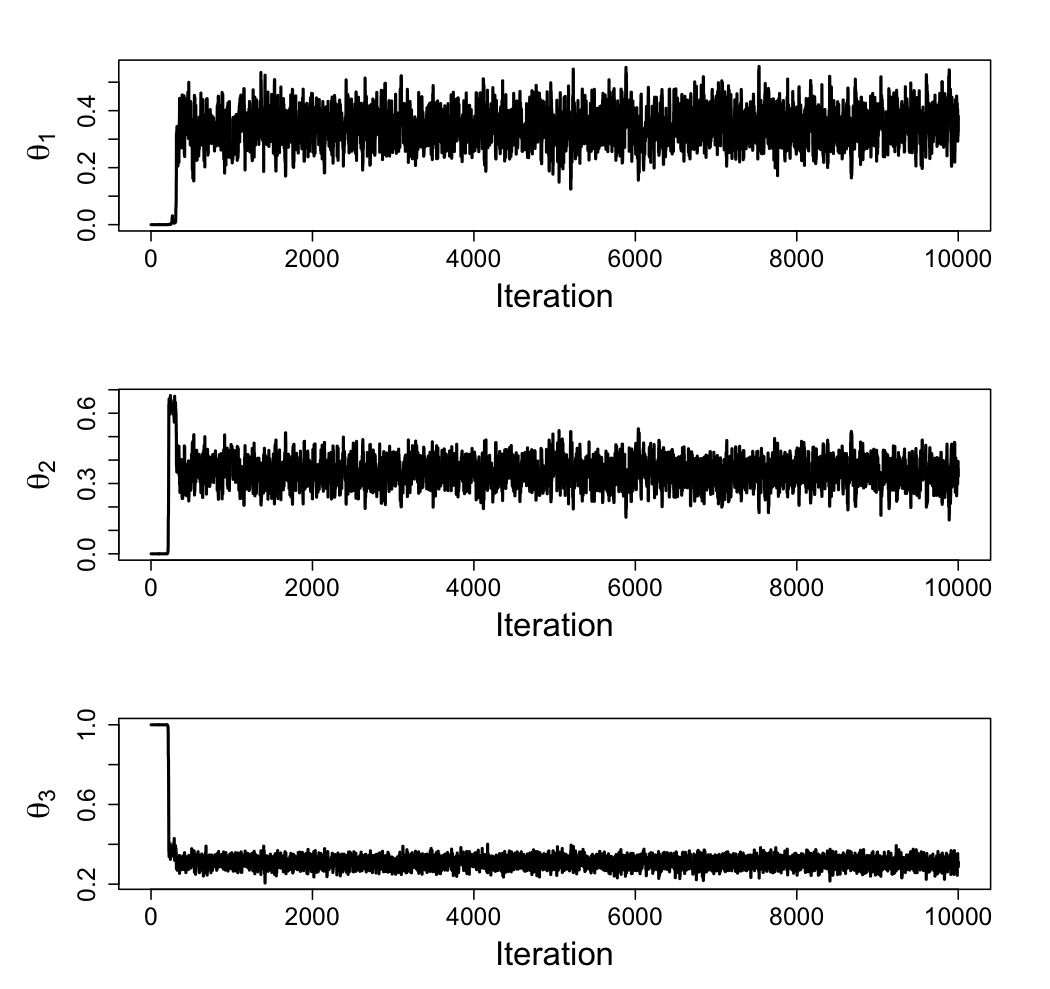}
	} 
	\subfloat[][SPInS joint sampler \label{fig:msn-spins-mult-tr}]{%
		\includegraphics[width=0.5\textwidth]{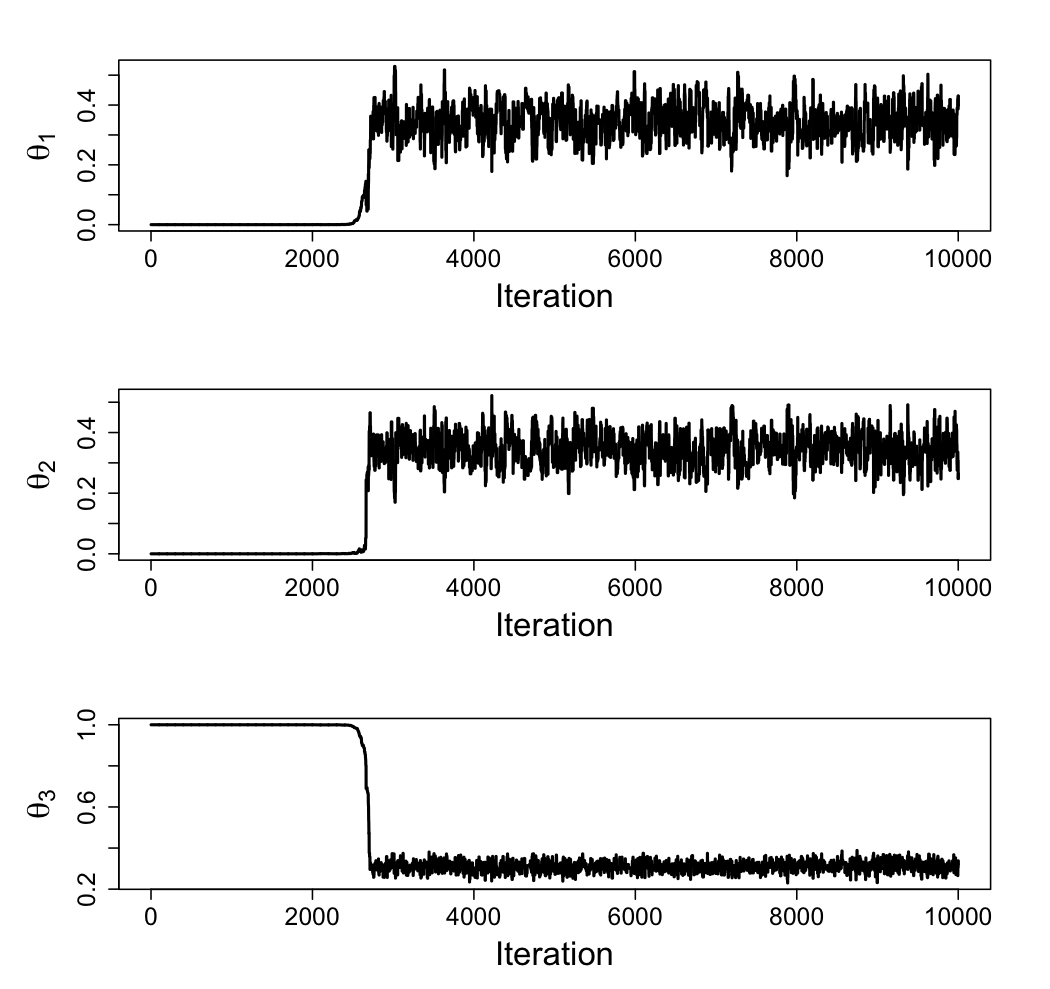}
	} \\
	\subfloat[][SALT sampler \label{fig:msn-salt-tr}]{%
		\includegraphics[width=0.5\textwidth]{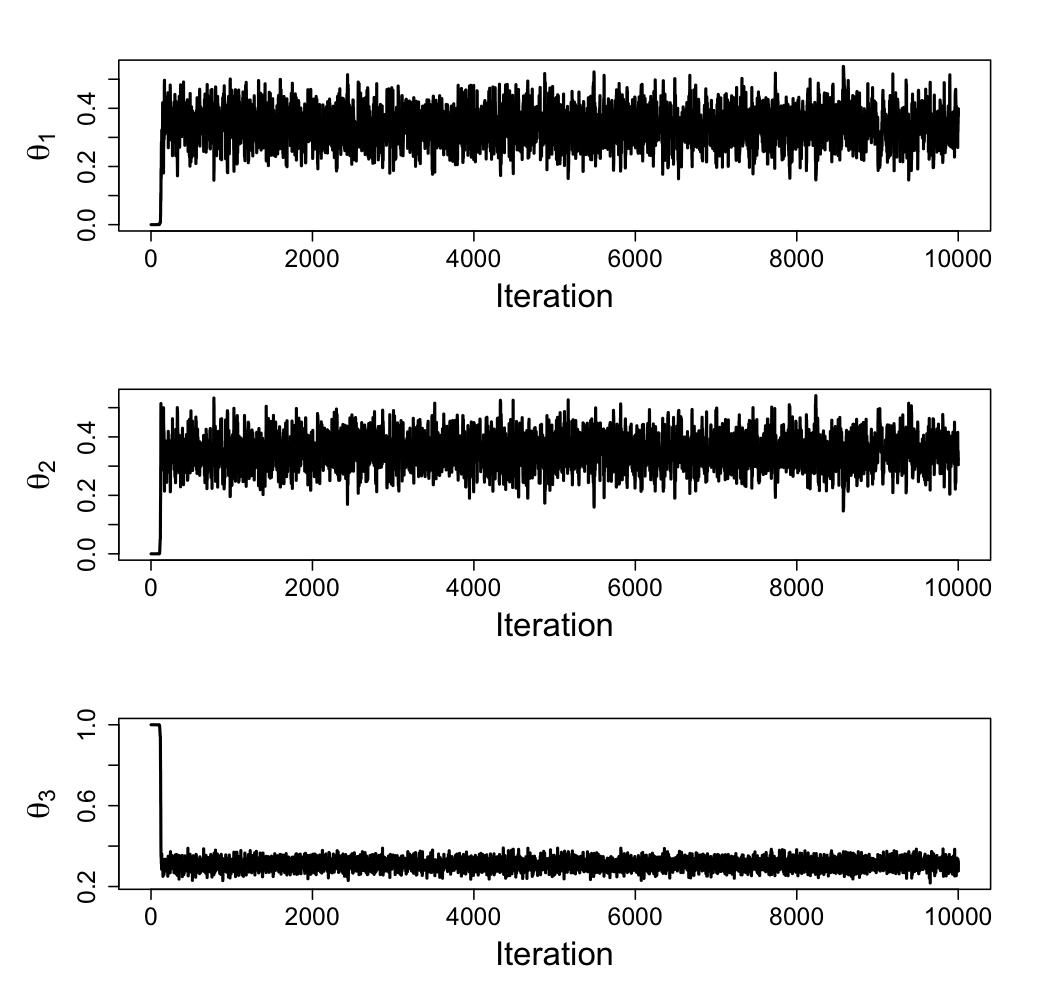}
	} 
	\subfloat[][Adaptive Dirichlet sampler \label{fig:msn-dirichlet-tr}]{%
		\includegraphics[width=0.5\textwidth]{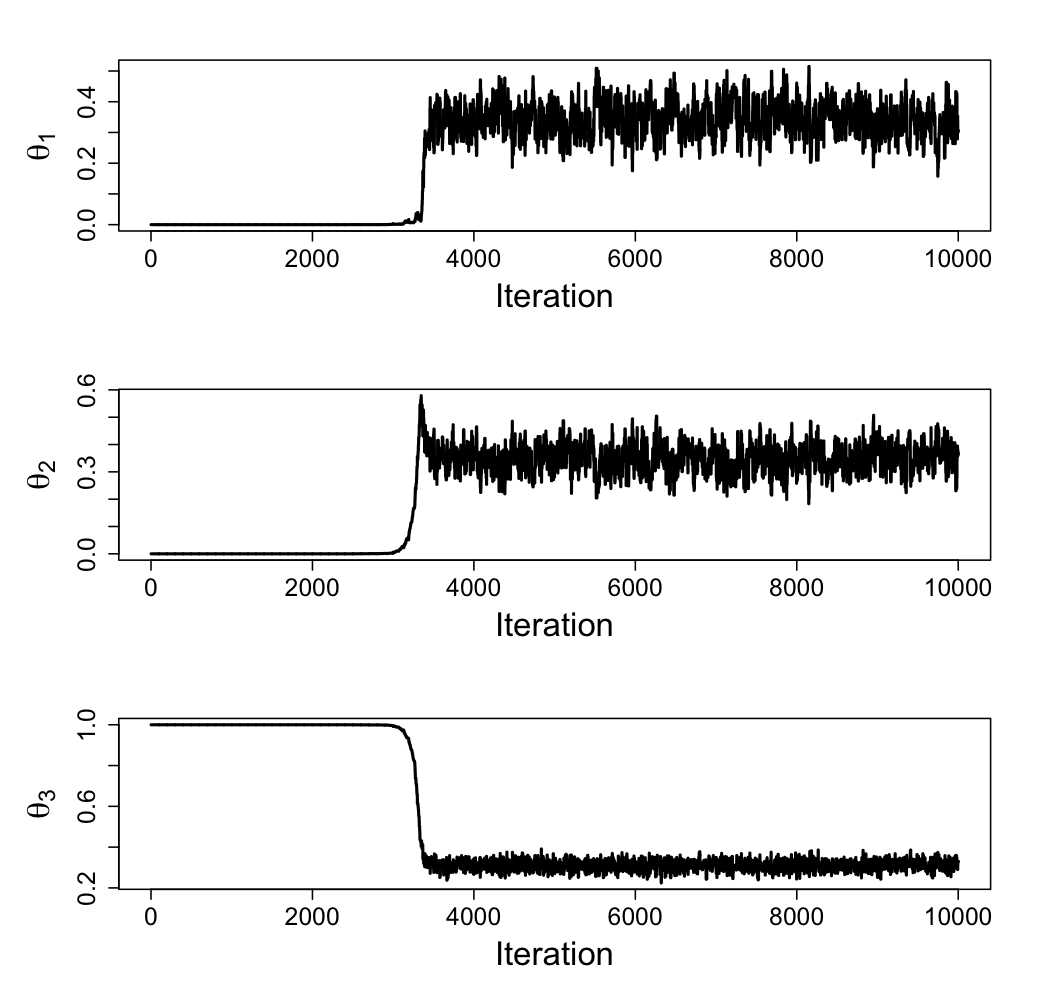}
	}	
\end{figure}

\noindent Fig. \ref{fig:pos-msn} shows the posterior sample obtained from each of the samplers on the $2-$simplex projected down in two dimensions. The Dirichlet and multivariate SPInS samplers appear to take a longer time to navigate the corner of the simplex in comparison to the componentwise SPInS and SALT samplers. This is confirmed by observing the corresponding trace plots in Fig. \ref{fig:pos-msn-tr}.

\subsubsection{Multiplicative Uncorrelated Error}
\label{sec:multerror}
\begin{figure}[h]
	\centering
	\caption{Posterior sample for multiplicative Gaussian noise}\label{fig:pos-mult-norm}
	\subfloat[][SPInS componentwise sampler \label{fig:mult-norm-spins-comp}]{%
		\includegraphics[width=0.5\textwidth]{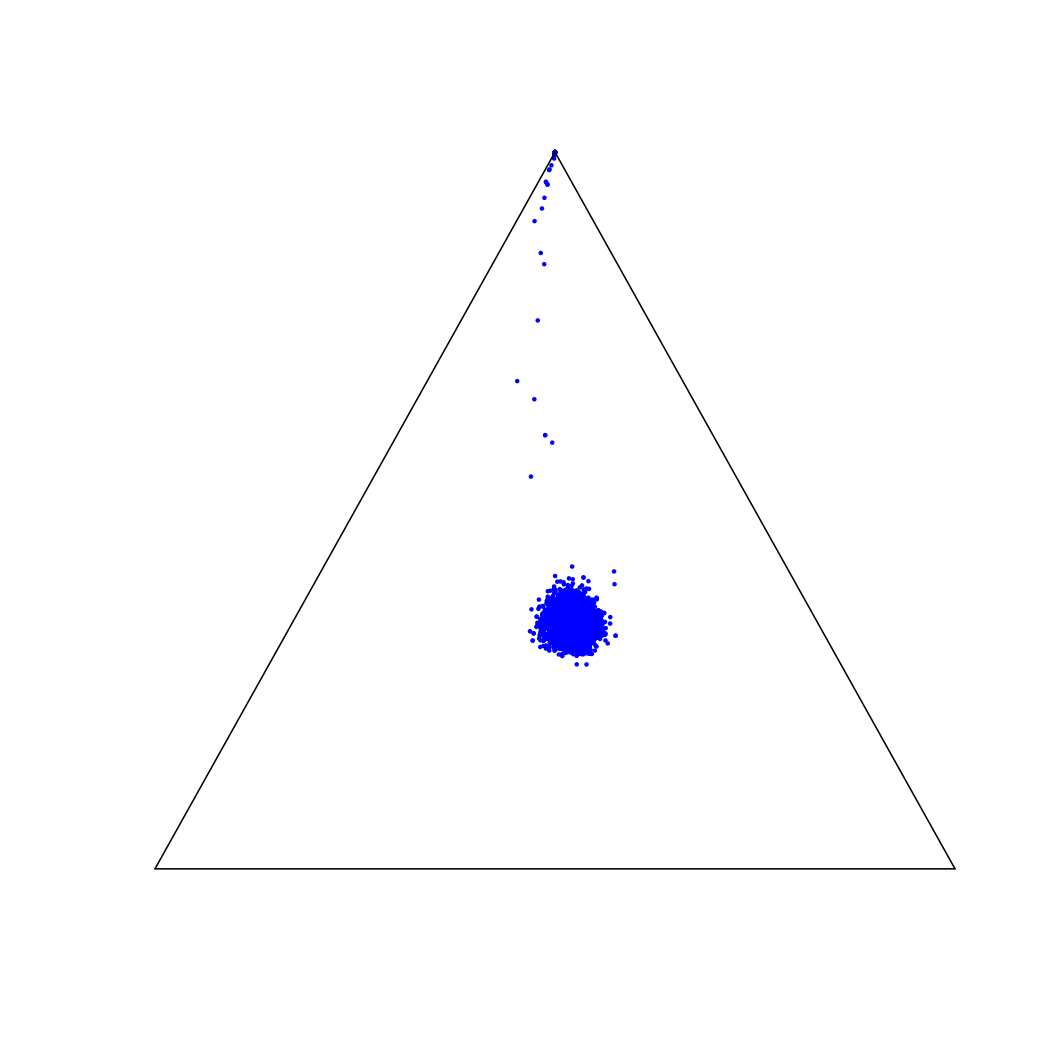}
	} 
	\subfloat[][SPInS joint sampler \label{fig:mult-norm-spins-mult}]{%
		\includegraphics[width=0.5\textwidth]{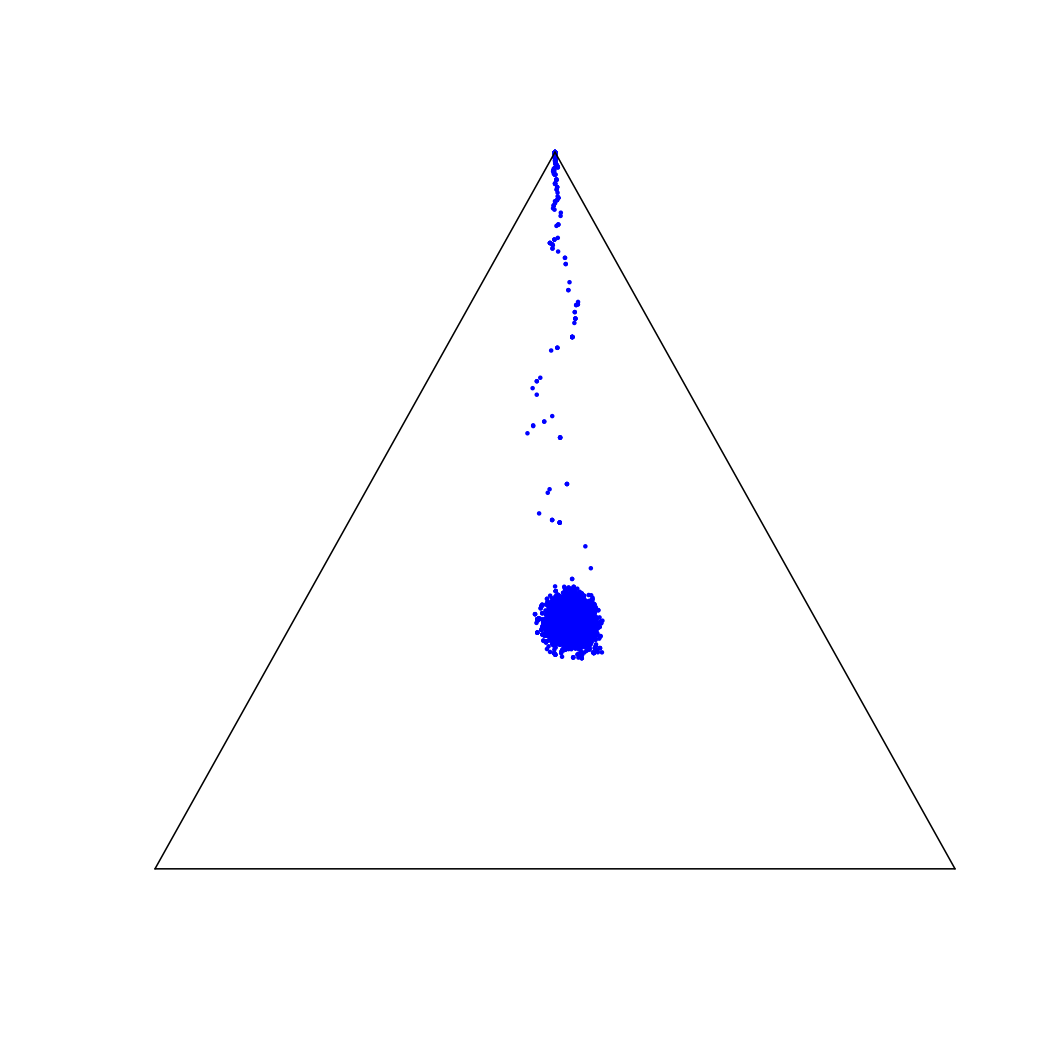}
	} \\
	\subfloat[][SALT sampler \label{fig:mult-norm-salt}]{%
		\includegraphics[width=0.5\textwidth]{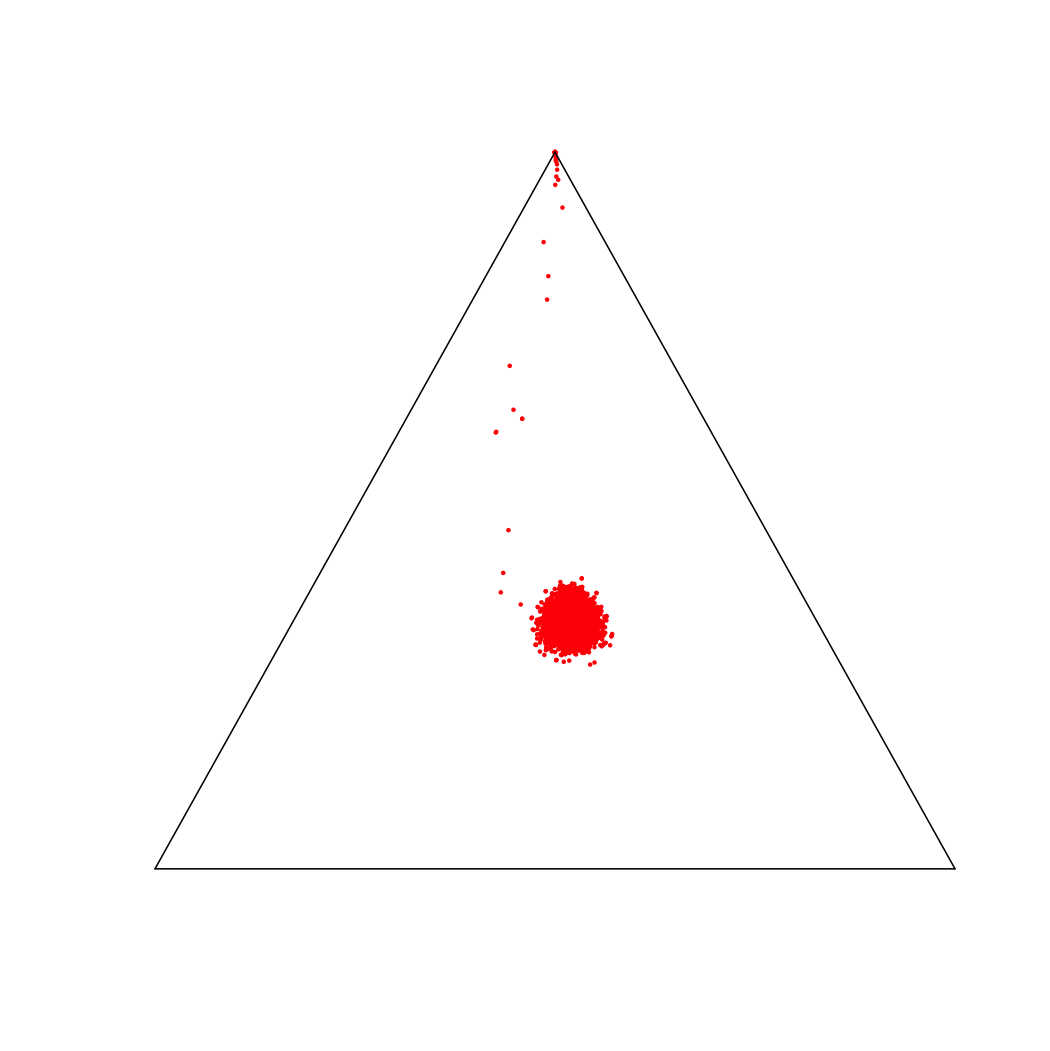}
	} 
	\subfloat[][Adaptive Dirichlet sampler \label{fig:mult-norm-dirichlet}]{%
			\includegraphics[width=0.5\textwidth]{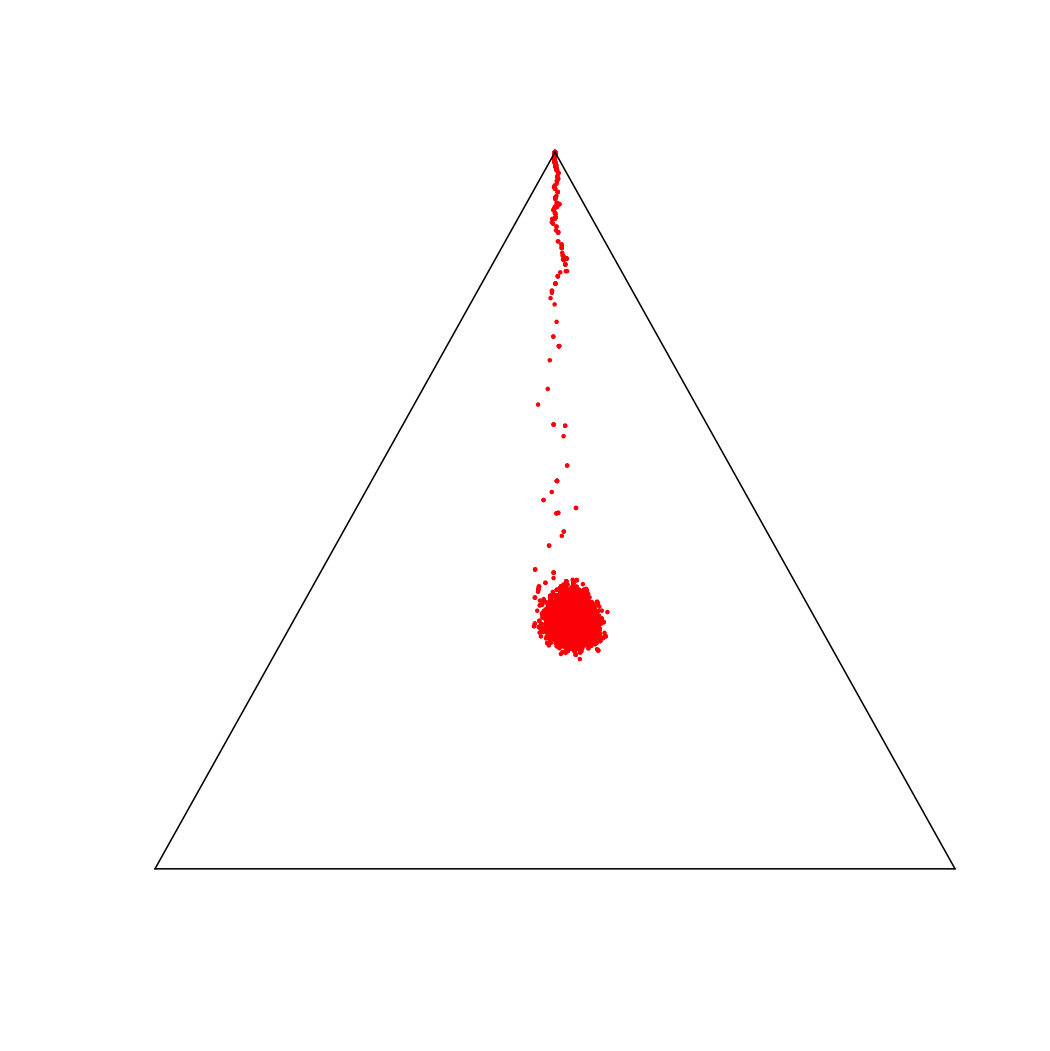}
	}
\end{figure}

\begin{figure}[h]
	\centering
	\caption{Trace plots for multiplicative Gaussian noise}\label{fig:pos-mult-norm-tr}
	\subfloat[][SPInS componentwise sampler \label{fig:mult-norm-spins-comp-tr}]{%
		\includegraphics[width=0.5\textwidth]{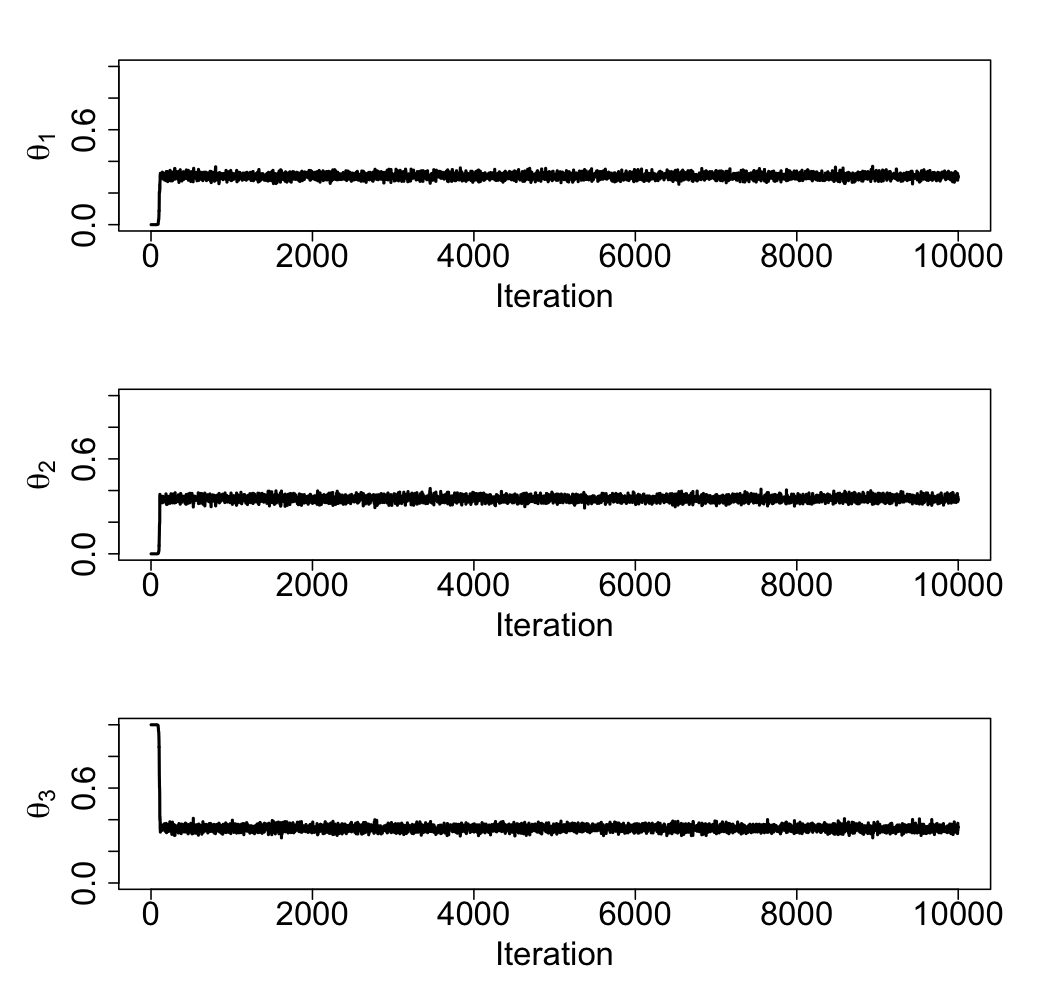}
	} 
	\subfloat[][SPInS joint sampler \label{fig:mult-norm-spins-mult-tr}]{%
		\includegraphics[width=0.5\textwidth]{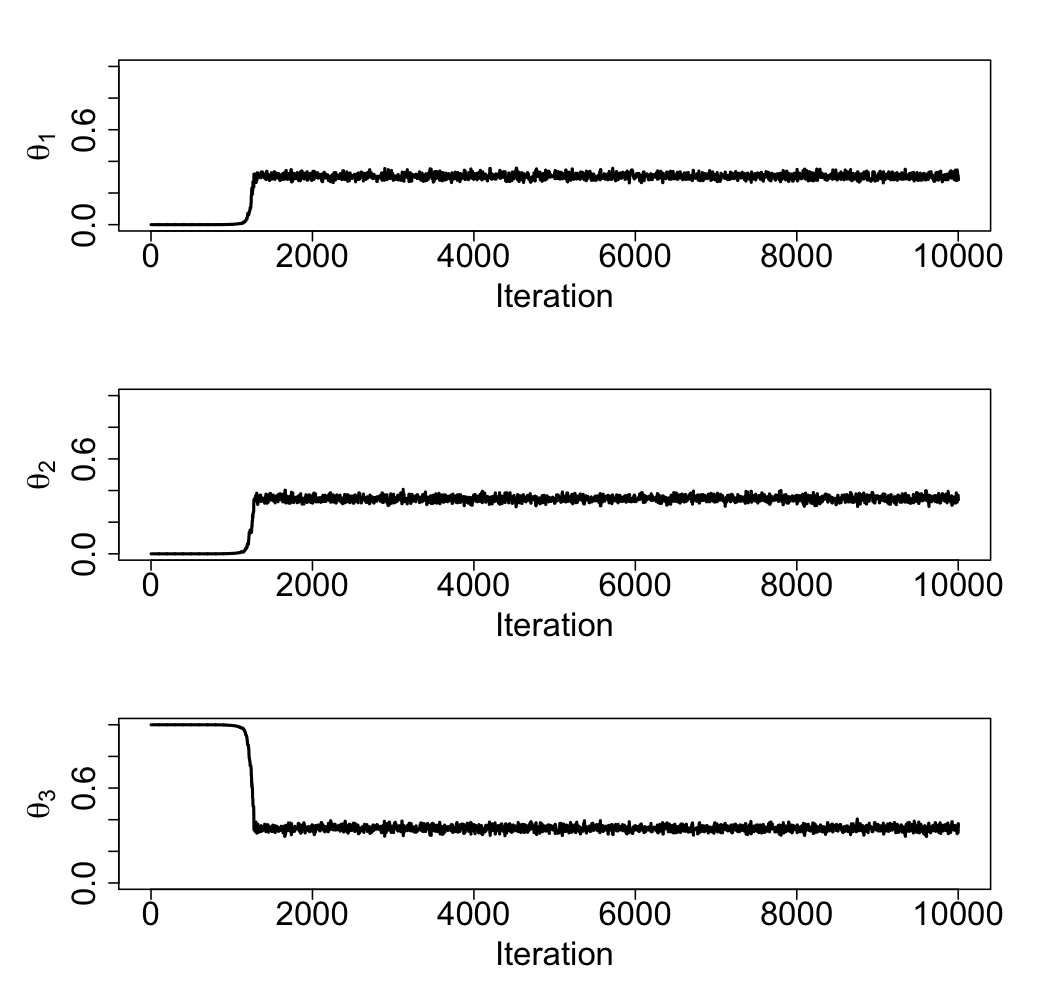}
	} \\
	\subfloat[][SALT sampler \label{fig:mult-norm-salt-tr}]{%
		\includegraphics[width=0.5\textwidth]{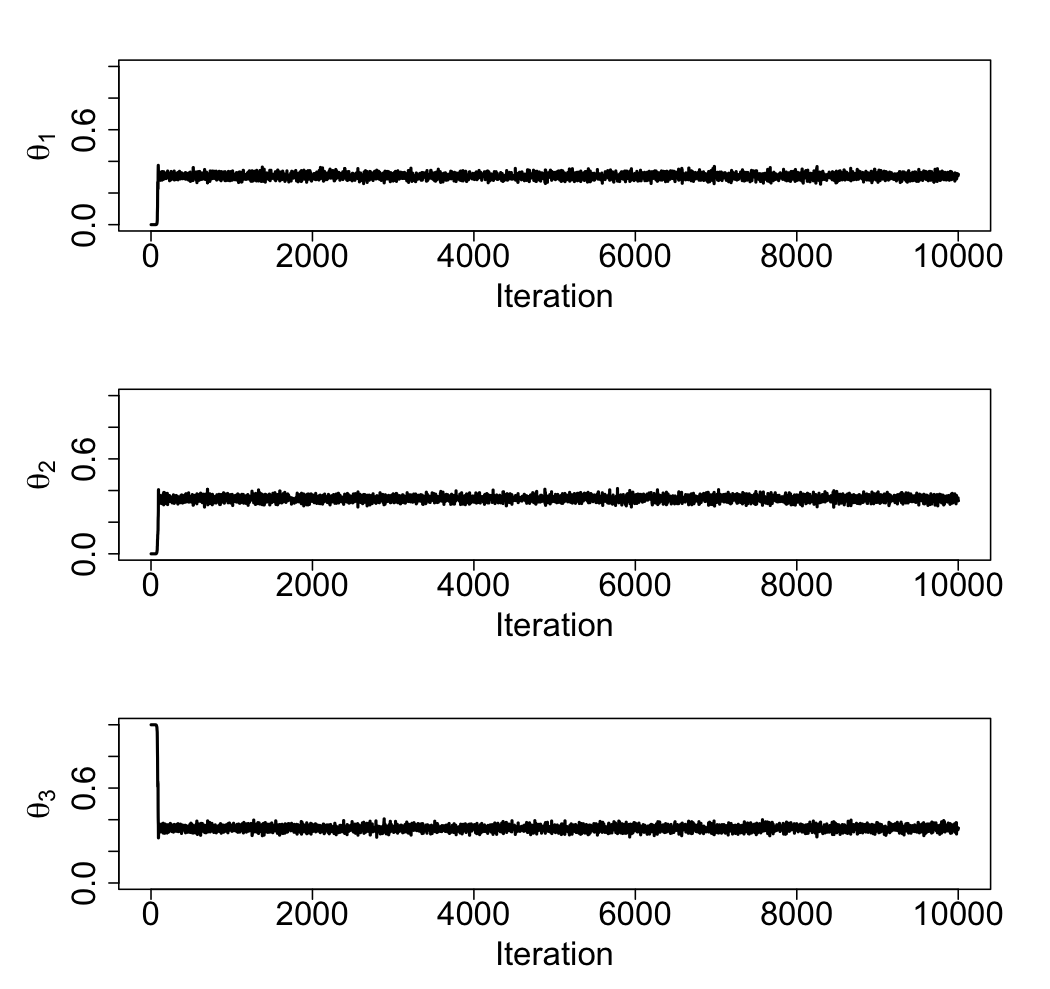}
	} 
	\subfloat[][Adaptive Dirichlet sampler \label{fig:mult-norm-dirichlet-tr}]{%
			\includegraphics[width=0.5\textwidth]{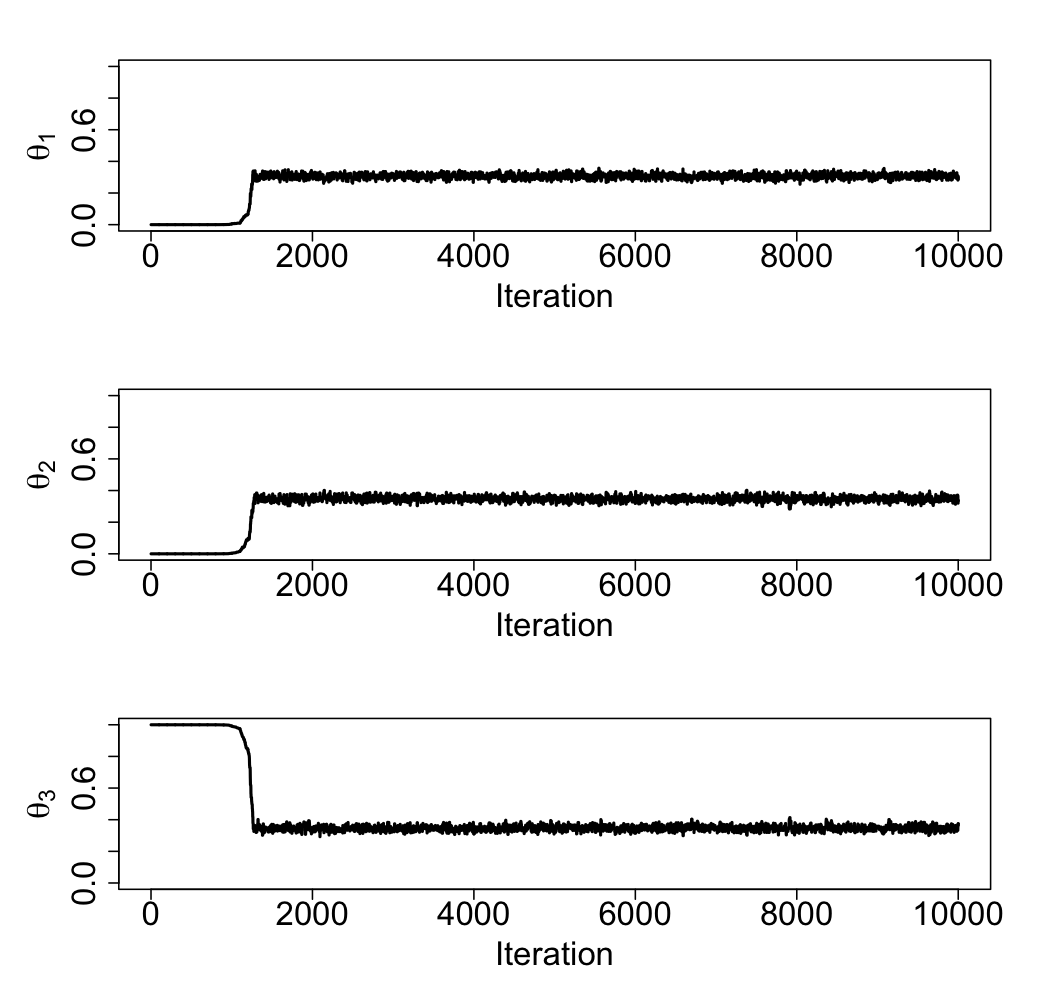}
	}
\end{figure}

\noindent To add a sense of generalization to the previous example, another dataset of 1000 observations was generated using componentwise multiplicative noise. That is, noisy signals $\mathbf{y} = \boldsymbol{\theta} \odot \boldsymbol{\epsilon}$ such that $\epsilon_j \stackrel{iid}{\sim} N(0,10^2)$ are observed. Here, $\odot$ represents componentwise multiplication of two vectors. The scaling parameters for componentwise SPInS, joint SPInS, SALT, and adaptive Dirichlet samplers were determined to be $4, 6, 0.3,$ and $50$ respectively. The samplers were initialized in the same manner as in the previous example. Figs. \ref{fig:pos-mult-norm} and \ref{fig:pos-mult-norm-tr} show the posterior samples on the $2-$simplex and their corresponding trace plots respectively, and a similar trend in terms of speed of convergence is observed in this example as well.

\subsubsection{Comments on simplex sampling}
\label{sec:Disc}
The efficiency of the multivariate SPInS sampler was found to be comparable to the adaptive Dirichlet sampler. Furthermore, the superiority of the (componentwise) SALT sampler, as shown in the examples above and previously explored in \cite{director2017efficient}, is shared by the componentwise SPInS sampler. Note that the Dirichlet distribution and the underlying $\mathrm{logit}$ transformation (of SALT) are natively limited to the simplex and the unit interval respectively. Although the SPInS algorithm does not have any obvious canonical relationship to the simplex, it still performs as well as other strategies that are tailored for the simplex. This makes the outlook for SPInS, a general strategy for sampling in constrained domains, extremely promising. To consolidate this narrative, an implementation of SPInS beyond the simplex is explored in Sections \ref{sec:nsphere} and \ref{sec:hypercube}. \\

\noindent Next, it is possible to explain why the performance of the componentwise SPInS sampler was remarkably close to that of the SALT sampler. This was chiefly because the proposal densities of the two methods ended up being similar even though two completely different transformations (inversion in a sphere vs. $\mathrm{logit}$) were employed. Specifically, compare the SPInS-Gaussian proposal density in (\ref{propdens}) to the SALT-Gaussian density

\begin{equation}
\label{SALTpropdens}
\begin{split}
q_{SALT}(\theta^*_i|\theta_i)  = & \dfrac{1}{\sqrt{2 \pi\ h_i^2}}\ \exp\left\{ \dfrac{-(\mathrm{logit}(\theta_i^*) - \mathrm{logit}(\theta_i))^2}{2\ h_i^2 }\right\} \cdot \\ & \dfrac{1}{\theta_i^* (1-\theta_i^*)}.
\end{split}
\end{equation}

\noindent There are two notable differences with the most significant being the difference in the Jacobian terms: $$\begin{cases}  \frac{1}{(\theta^*_i)^2} & \text{if}\ \theta_i \leq \frac{1}{2} \\ \frac{1}{(\theta^*_i-1)^2} & \text{if}\ \theta_i > \frac{1}{2} \end{cases}\quad  vs.\quad \dfrac{1}{\theta_i^* (1-\theta_i^*)}.$$ Yet, these two functions exhibit very similar behaviors on their domain (0,1). Note that they both explode to infinity at the endpoints and are equal to $4$ when $\theta_i^* = \frac{1}{2}$. The Jacobian terms, here, reshape the Gaussian distribution to adapt to the geometry of the simplex. It could be argued that the Jacobian for SPInS goes to infinity faster than SALT, but the choice of the Gaussian distribution dampens the effect of that explosion near 0 and 1. This property may be used as an advantage for some class of likelihood functions, but may require the employment of a proposal distribution other than the Gaussian (e.g. Uniform) where the effect of the Jacobian is more pronounced. Although interesting, this endeavor is considered beyond the scope of the work presented here.\\

\noindent The second difference is in the scale parameter of the Gaussian part of the proposal density. Since SALT maps the parameter space to the entire interval $(-\infty, \infty)$, the sampler is able to maintain the symmetric nature of the Gaussian distribution (on the image of the unit interval) by selecting a single common value of the scale parameter, $h$. In the case of the SPInS sampler, on the other hand, $q(\delta^*_i|\delta_i)$ and $q(\delta_i|\delta_i^*)$ have different scale parameters ($\frac{\eta_i}{d}$ and $\frac{\eta_i^*}{d}$ respectively). It is important to note that this symmetry is still numerically maintained $\left(\frac{q(\delta^*_i|\delta_i)}{q(\delta_i|\delta_i^*)} \approx 1\right)$ when $\delta_i \approx \delta_i^*$, which is typically the case once the Markov Chain converges or when a large enough value for $d$ is chosen. \\

\subsection{An application}
\label{sec:application}
This section explores the use of the SPInS sampler on a problem arising in neuroimaging. The ball-and-stick model (\ref{ballstick}), as proposed by \cite{behrens2007probabilistic} attempts to model signals obtained with diffusion magnetic resonance imaging (dMRI) within a given volumetric pixel (voxel) containing $K$-nerves. Voxels are 3-dimensional, typically cubic, analogs of rectangular pixels in 2-dimensional images. 

\begin{equation}
\label{ballstick}
\mu_i = S_0 \left[ f_0 e^{-b \lambda} + \displaystyle \sum_{j=1}^K f_j e^{-b \lambda \textbf{r}_i^t g(\theta_j,\phi_j) \textbf{r}_i }\right], i = 1,2,\dots,n.
\end{equation}
Here, $S_0$ represents the baseline signal with no diffusion gradient, $\textbf{r}_i$ is the direction of the $i^{th}$ diffusion gradient, $b_i$ is an experimentally set b-value for the $i^{th}$ signal, $\lambda$ is the apparent diffusivity, $\boldsymbol{f} = (f_0,f_1,\dots,f_K) $ is a vector of volume-fractions, $(\theta_j,\phi_j)$ represent the elevation and azimuthal angles respectively of the principal diffusion direction of the $j^{th}$ nerve fiber, $g(\cdot,\cdot)$ is a matrix that rotates around the elevation and azimuth angles, $\mu_i$ is the expected $i^{th}$ dampened diffusion signal, and $n$ represents the number of diffusion signals obtained. In the given model with $K$-nerves, $\left\{S_0, d, \boldsymbol{f}, (\boldsymbol{\theta},\boldsymbol{\phi}),\sigma^2\right\}$ are unknown parameters. \\

\noindent The observed dMRI signals are considered to be noisy versions of the expected signal in (\ref{ballstick}), and are obtained by adding Rician noise as shown in (\ref{rice}). The noise structure is chosen to ensure that the corrupted signals are indeed non-negative. For dMRI data, the signal-to-noise ratio (SNR) is defined as $\frac{S_0}{\sigma}$.

\begin{equation}
\label{rice}
S_i = \sqrt{\left(\mu_i + \epsilon_{i,1}\right )^2+\epsilon_{i,2}^2}\ ,\  \epsilon_{i,l} \sim N(0,\sigma^2), l=1,2.
\end{equation}

\noindent The focus, here, is specifically on $\boldsymbol{f}$ because this vector holds the proportion of volume occupied by each component, and since all components are accounted for in the model, the condition $\displaystyle \sum_{j=0}^K f_j= 1 $ holds. This is where the utility of the SPInS sampler is obvious.\\

\begin{figure}[h]
	\caption{A run of the SPInS sampler in the three nerve case of the ball-and-stick model}
	\label{fig:applicationrun}
	\centering
	\includegraphics[scale = 0.6]{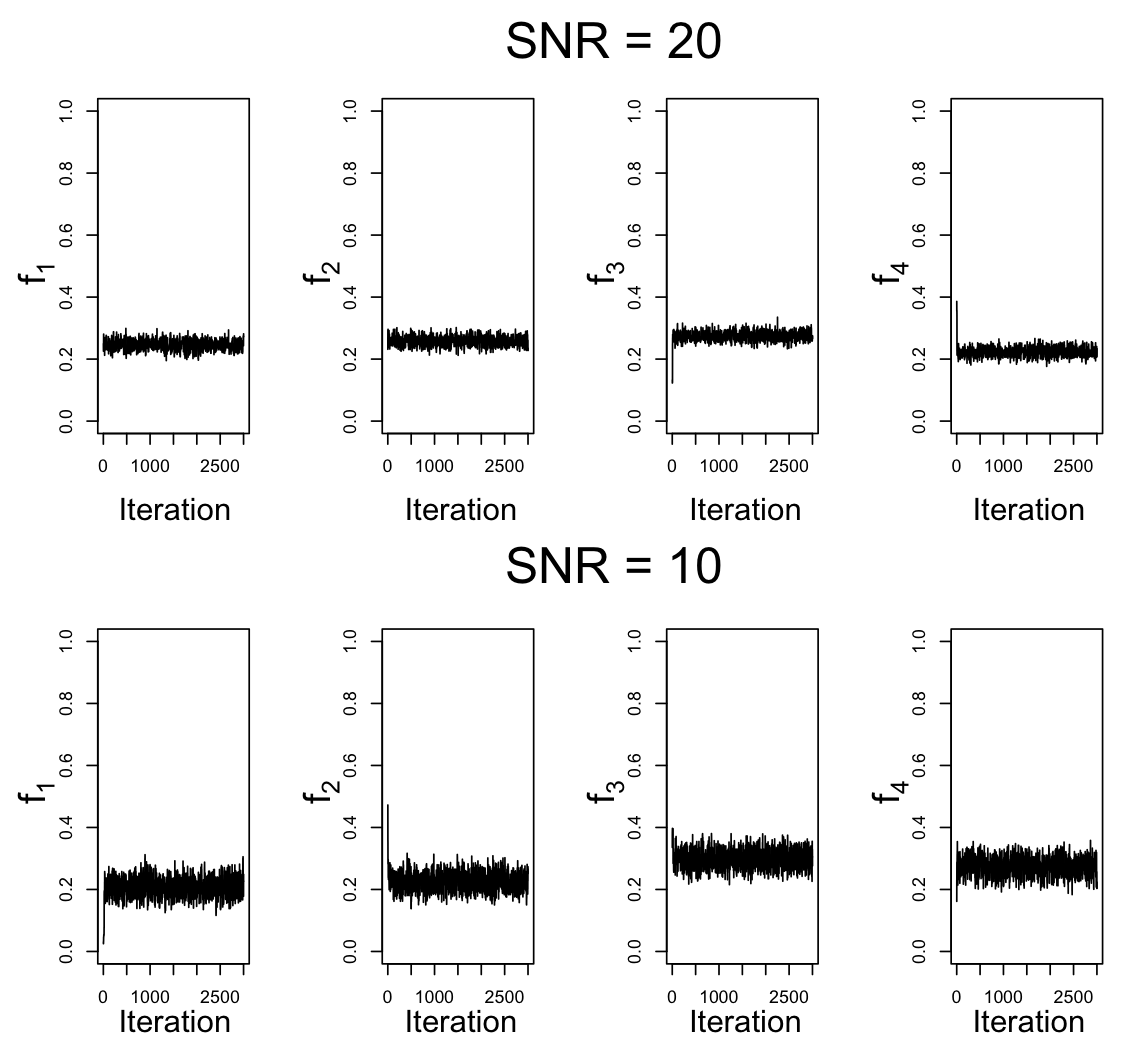}
\end{figure}

\noindent A dataset with $n=64$ observations is generated with $K=3$ at SNR = 20 and 10. The choice of dMRI parameters and SNRs are industry standard values used for model validation \cite{daducci2014quantitative}. All parameters other than $\boldsymbol{f}$ are assumed to be known. Furthermore, it is assumed that the data has additive Gaussian noise (instead of Rician noise used in the generating model) to exemplify the robustness of the technique. In the two datasets, the true value of $ \boldsymbol{f}$ was chosen to be $(0.25,0.25,0.25,0.25)$. A single run of the SPInS sampler (3,000 iterations) is presented for both SNRs in Fig. \ref{fig:applicationrun}. At SNR = 10, the posterior samples showed more variability as expected. $d$ was set to 3 in both these cases to maintain good mixing.

\section{SPInS on other domains}
\label{sec:beyond}
The charm of the SPInS procedure lies in the fact that the principles detailed in Section \ref{sec:simplex} can be generalized to sample other constrained parameter spaces. Two additional implementations of the joint SPInS strategy are presented in this section. Although not shown, it is argued that the componentwise strategy may be replicated as well. For brevity, the SPInS procedure is not reiterated in these scenarios, but it should be noted that there are two points of difference. The first involves obtaining the projection of the current value onto the boundary of the region to be sampled and the second involves obtaining an image of this region when inverted in a sphere for the purpose of finding $\eta$. The relevant details are provided in Appendix \ref{sec:app-A}. It is also worth mentioning that $T$ can be shown to be one-to-one by following the same proof as presented in the case of the simplex in Appendix  \ref{sec:app-B}.

\subsection{Sampling in a sector of an $n$-sphere}
\label{sec:nsphere}
Consider the domain of the parameters to be a unit $n$-sphere in the nonnegative orthant of $\mathbb{R}^n$. In two dimensions this is equivalent to a quarter unit circle in the first quadrant. In other words, the parameter space is:

\begin{equation}
\label{eq:nsphere}
\mathcal{B} := \left\{ \boldsymbol{\theta} = (\theta_1,\theta_2,\dots,\theta_n) \in \mathbb{R}^n\ |\ \theta_i \geq 0, i = 1,2,\dots,n, \displaystyle \sum_{i=1}^n \theta_i^2 \leq 1\right\}.
\end{equation}

\noindent Unlike the simplex, there is no requirement of projecting $\mathcal{B}$ to $\mathbb{R}^{n-1}$. Furthermore, $\mathcal{B}$ is made up of the same half planes as $\mathcal{S_-}$ except for the surface of the $n$-sphere. \\

\noindent A dataset comprising 1000 observations with additive Gaussian noise is generated. I.e., $\mathbf{y} = \boldsymbol{\theta} + \boldsymbol{\epsilon}$ is observed such that $\epsilon_j \stackrel{iid}{\sim} N(0,1)$. The true value of $\boldsymbol{\theta}$ is set to $(0.4, 0.5, 0.6)$. The performance of the SPInS sampler is compared to a classical MH-sampler with uniform proposal on $\mathcal{B}$. Both samplers are initialized at a randomly selected point in the interior of $\mathcal{B}$. A reasonable choice of $d$ for the SPInS sampler was determined to be 3.\\

\begin{figure}[h]
	\caption{Posterior sample in a sector of the sphere using SPInS proposal (left) and Uniform proposal (right)}
	\label{fig:SphSec}
	\centering
	\includegraphics[scale = 0.25]{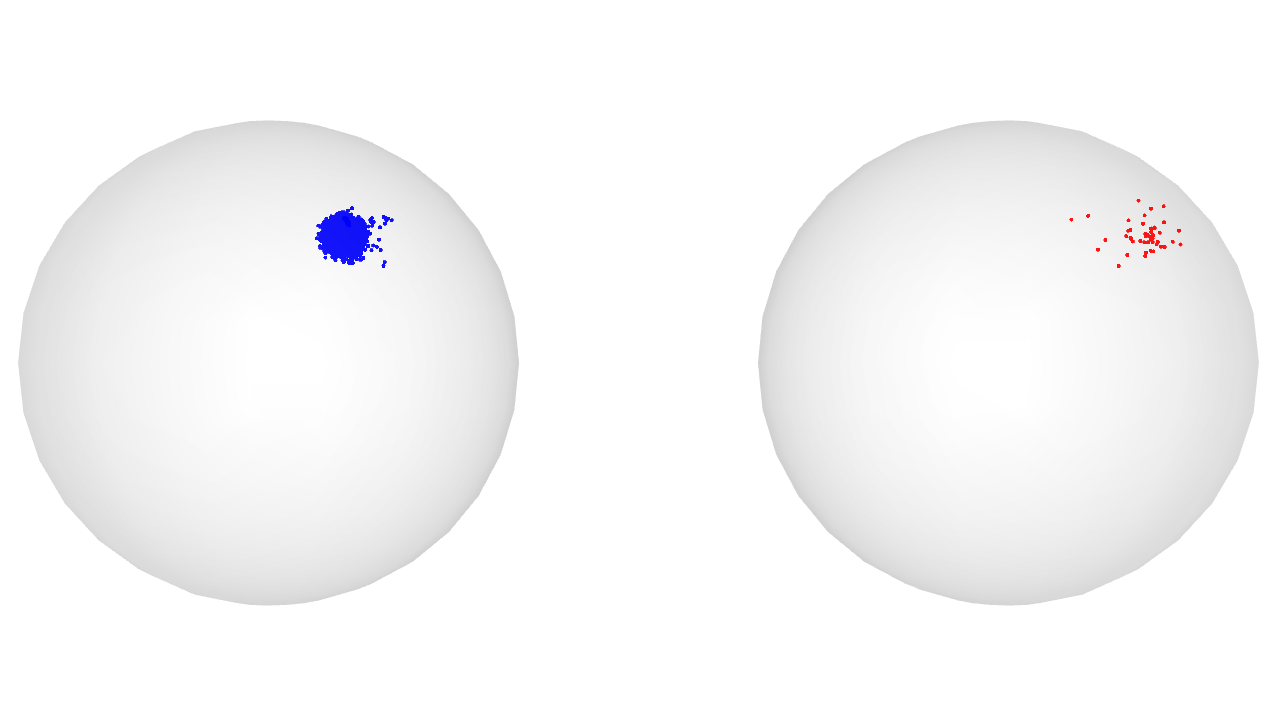}
\end{figure}

\begin{figure}[h]
	\centering
	\caption{Trace plots for additive Gaussian noise in a sector of a sphere}\label{fig:SphSec-tr}
	\subfloat[][SPInS proposal \label{fig:SphSec-spins-tr}]{%
		\includegraphics[width=0.6\textwidth]{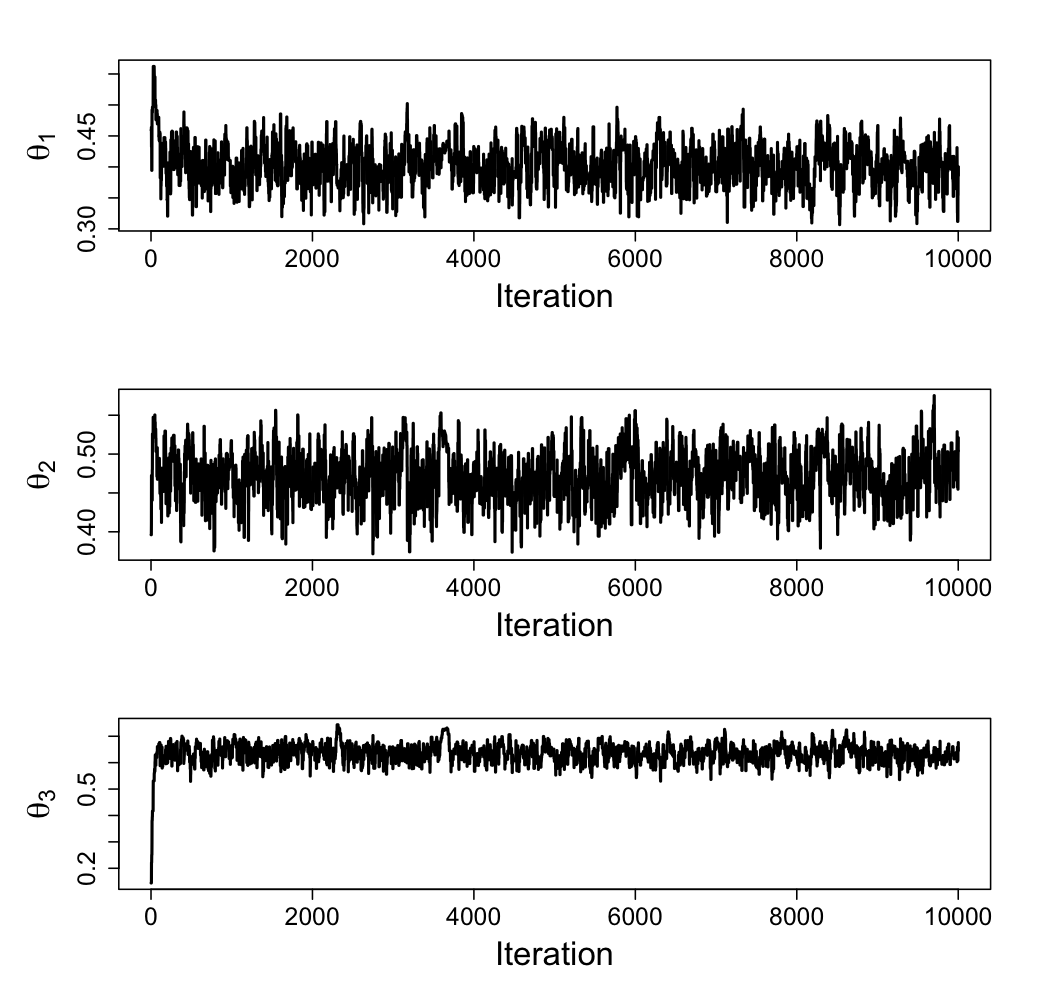}
	} \\
	\subfloat[][Uniform proposal \label{fig:SphSec-unif-tr}]{%
		\includegraphics[width=0.6\textwidth]{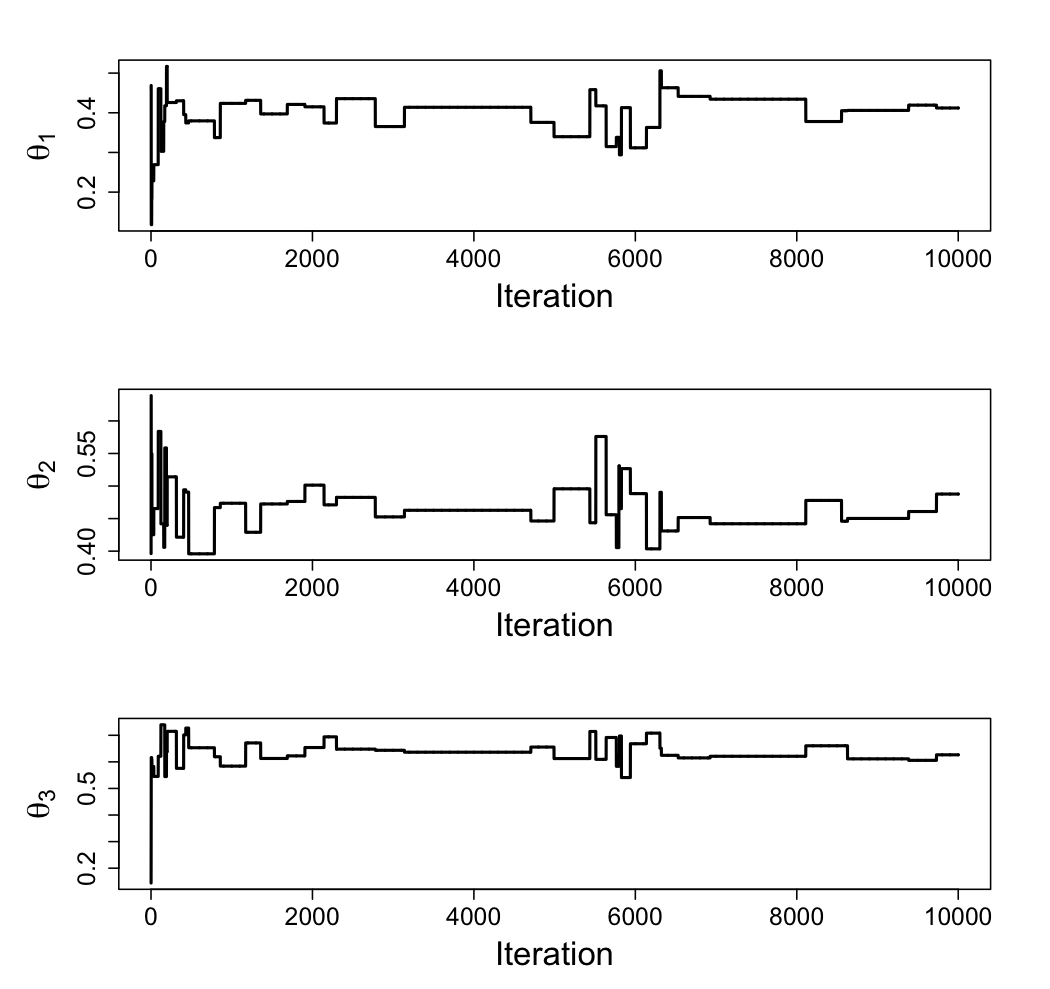}
	}
\end{figure}

\noindent Fig. \ref{fig:SphSec} presents the posterior samples for a run of each algorithm in the positive sector of the unit sphere. Each sampler was run for 10,000 iterations and it is easily seen that the SPInS posterior sample is extremely dense in comparison to the uniform proposal. This discrepancy of mixing rates between the two samplers is again visually emphasized in the trace plots presented in Fig. \ref{fig:SphSec-tr}. The key tenet behind this experiment is that when a choice of a naturally adaptable distribution is not obvious, SPInS may be an effective alternative. 

\subsection{Sampling in a hypercube}
\label{sec:hypercube}
The final domain explored for sampling in this work is a hypercube with one vertex at the origin and edge length $a>0$. That is, sampling needs to be performed in the following set:
\begin{equation}
\label{eq:hypercube}
\mathcal{H} := \left\{ \boldsymbol{\theta} = (\theta_1,\theta_2,\dots,\theta_n) \in \mathbb{R}^n\ |\ a \geq \theta_i \geq 0, i = 1,2,\dots,n, \right\}.
\end{equation}
\noindent Just like on $\mathcal{B}$, SPInS sampling can be performed natively on $\mathcal{H}$. \\

\begin{figure}[h]
	\hspace*{-5in}
	\caption{Trace plots for additive Gaussian noise in 10-dimensional cube}\label{fig:cube-tr}
	\subfloat[][SPInS \label{fig:SphSec-spins-tr}]{%
		\includegraphics[width=0.5\textwidth]{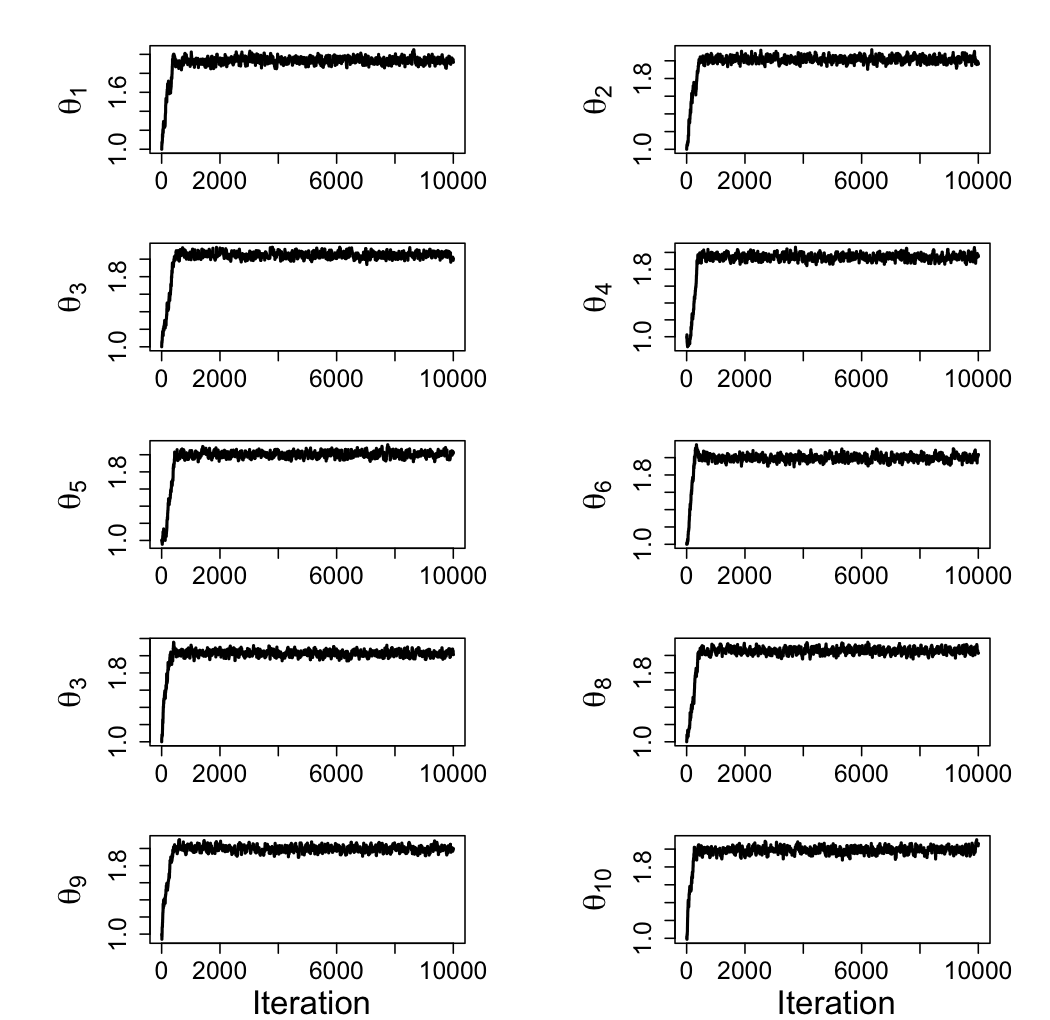}
	} \\
	\subfloat[][Uniform proposal joint update \label{fig:SphSec-unif-tr}]{%
		\includegraphics[width=0.5\textwidth]{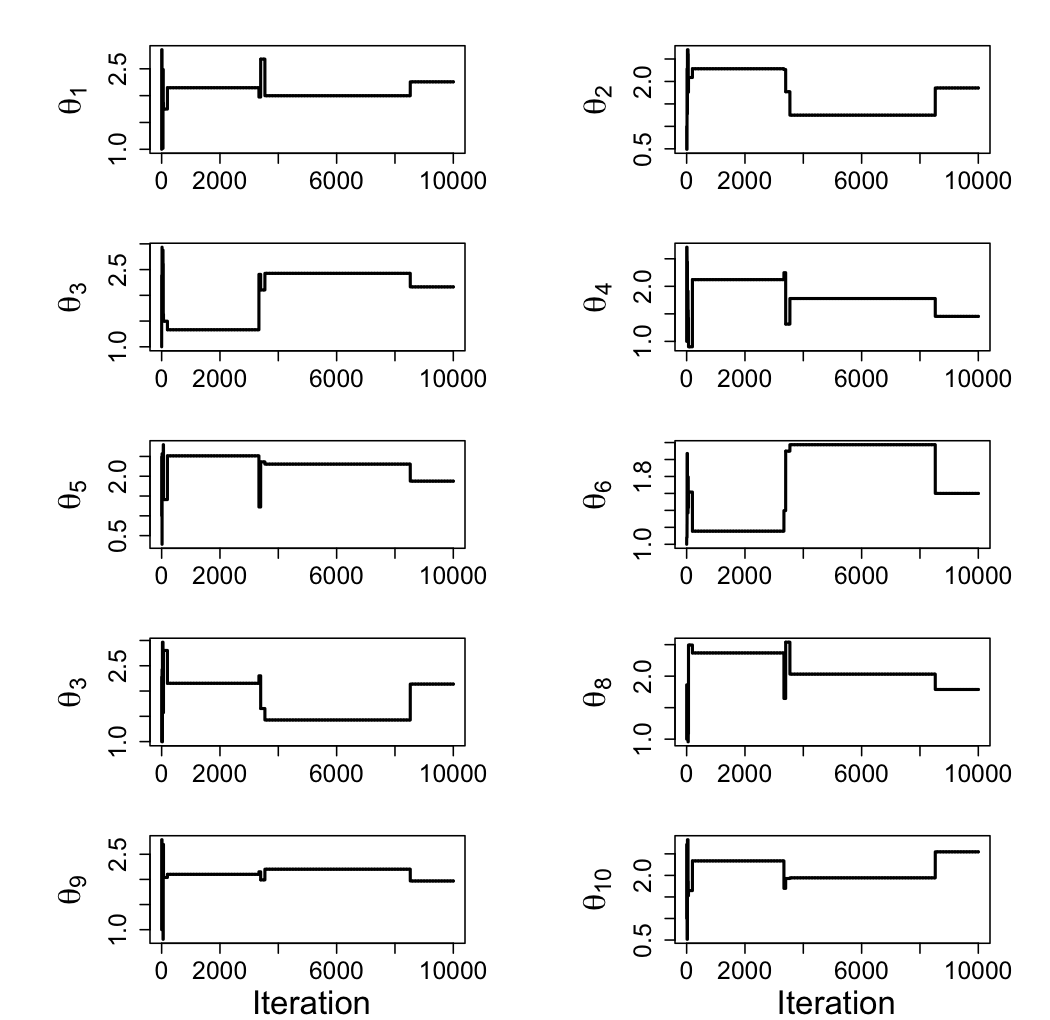}
	} 
	\subfloat[][Uniform proposal componentwise update \label{fig:SphSec-unif-tr}]{%
		\includegraphics[width=0.5\textwidth]{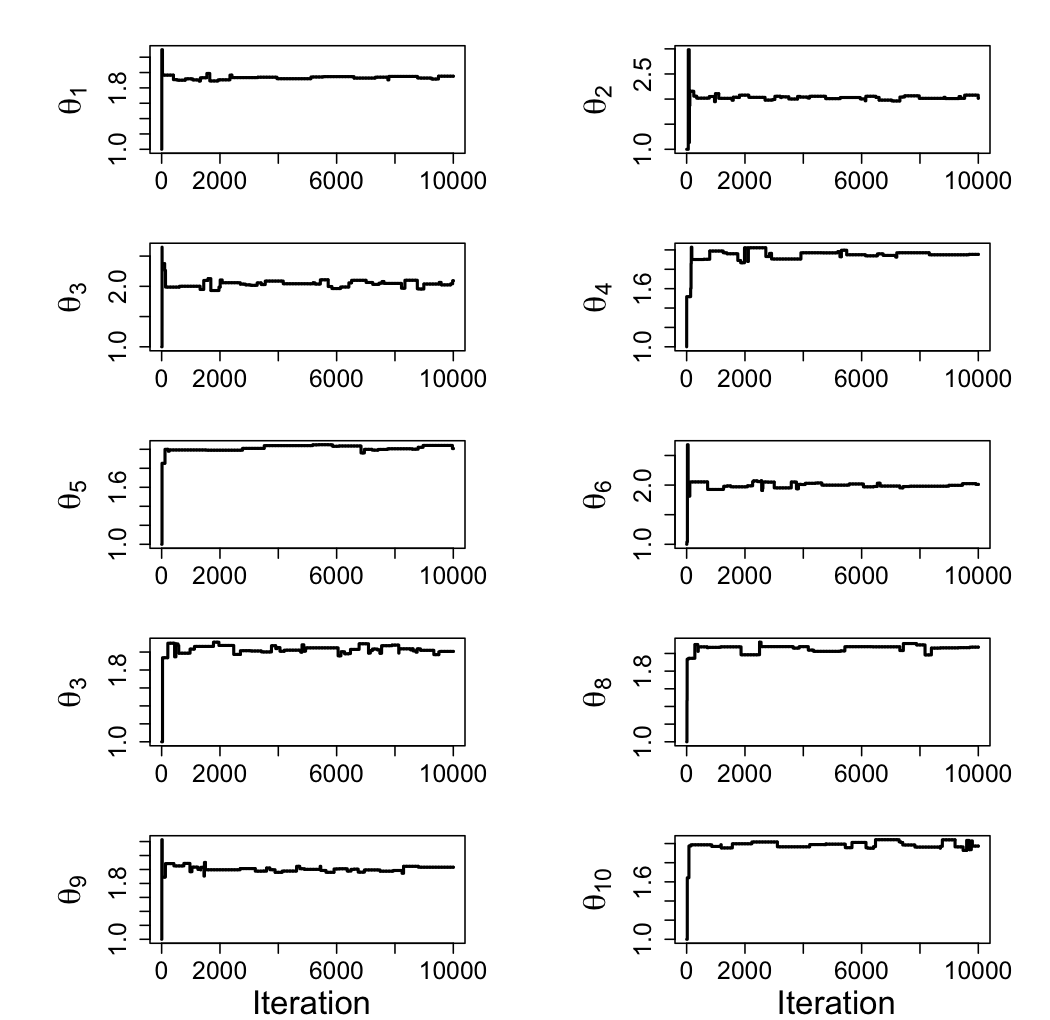}
	}
\end{figure}

\noindent In this example, a 10-dimensional hypercube of edge length 3 is chosen as the region of interest. A dataset with 1000 observations is generated with additive standard Gaussian noise $(N(0,1))$ with $\boldsymbol{\theta} = 2\cdot \boldsymbol{1}$, where $\boldsymbol{1}$ is a vector with every entry equal to 1. The SPInS sampler is compared to a multivariate uniform sampler and a componentwise uniform sampler. In the multivariate uniform sampler, a new value for each component is proposed using $Unif(0,3)$ and then either accepted or rejected. For the componentwise sampler, a new value is proposed for only a single component using $Unif(0,3)$ and then accepted or rejected. All three samplers are initialized at the point $\boldsymbol{1}$. The $d$ parameter is chosen to be 30. Fig. \ref{fig:cube-tr} presents the trace plots for the three samplers and it is evident that SPInS significantly outperforms both Uniform samplers.

\section{Conclusion and future work}
\label{sec:conc}
\noindent A novel scheme using inversion spheres that attempts to make constrained parameter spaces amenable to sampling is proposed in this work. The method was illustrated on the standard simplex by using a componentwise and joint MH-update strategy, and found to be competitive with the state of the art strategies. The appeal of the method relies on its generalizability to other constrained parameter spaces. Therefore, the method was exemplified in a sector of an $n$-sphere and within a hypercube. In all these experiments, a single choice of a scaling parameter was chosen for the proposal distribution. An integrated framework that utilizes pilot tuning for achieving faster convergence can be explored in the future. In addition, information during sampling may be used to get empirical estimates of the covariance matrix in the case of the multivariate sampler. Although there is plenty of literature on these aspects for MCMC methods in general, the fact that a new image of the constrained parameter space is created at every iteration poses a challenge.\\

\noindent Possibly the most attractive quality of the proposed work is the ability to handle varied constraints. Naturally, the efficacy of the method is related to obtaining a tractable image of the region of interest. Inversion in a sphere is relatively straightforward when inverting planes and spheres, but may be tedious otherwise. It would also be worth exploring the challenges of concave domains and test the performance of this method. Therefore, a thorough investigation on the limitations of SPInS may be worth undertaking.

\appendix

\bigskip
\begin{center}
	{\Large\bf{Appendix}}
\end{center}
\section{Inversion Procedure}
\label{sec:app-A}

\noindent In $ \mathbb{R}^n $, inversion in a sphere of radius $ r $ with center $ \boldsymbol{\alpha} = (\alpha_1,...,\alpha_n) $ is given by
\begin{align}
\begin{pmatrix}
x_1\\\vdots \\ x_n
\end{pmatrix} \mapsto \begin{pmatrix}
\alpha_1 + \frac{r^2(x_1-\alpha_1)}{(x_1-\alpha_1)^2 + \cdots + (x_n - \alpha_n)^2} \\ \vdots \\ \alpha_n + \frac{r^2(x_n-\alpha_n)}{(x_1-\alpha_1)^2 + \cdots + (x_n-\alpha_n)^2}
\end{pmatrix}.
\end{align}

\noindent Note that $ \mathcal{S}_- \subset \mathbb{R}^n $ is the intersection of $ n+1$ half spaces defined by the planes $\left\{x_j =0 \right\}_{j=1}^{n}$ (let's call these $ W_1,...,W_{n}$) and $\displaystyle \sum_{j=1}^n x_j = 1$ (called $W_{n+1}$). Given a point $ \boldsymbol{x} \in{\mathcal{S}}_- $, the center $ \boldsymbol{\alpha} $ of inversion will be given by the projection of $ \boldsymbol{x}$ onto the nearest plane $ W_j $. \\

\noindent The projection of $ \boldsymbol{x}$ onto the edges $ W_1,...,W_{n} $ is given by 
\begin{align}
\text{Proj}_{W_j}\boldsymbol{x} = (x_1,...,x_{j-1},0,x_{j+1},...,x_n),
\end{align}   
and the projection onto $ W_{n+1} $ which is defined by the equation $ x_1+\cdots + x_n = 1 $ is given by
\begin{align}
\text{Proj}_{W_{n+1}}\boldsymbol{x} = \frac{1}{n}((nI-M)\boldsymbol{x} + \boldsymbol{n}),
\end{align}
where $ I $ is the $ n \times n $ identity matrix, $ M = [1]_{n \times n} $, and $ \boldsymbol{n} = [1]_{1 \times n}$.\\

\noindent Note that spherical inversion sends spheres to spheres, and since a plane is a sphere containing the point at infinity, $ T(W_j) $ are also spheres with finite radius when $ \boldsymbol{\alpha} \not \in{W_j} $. We wish to find the distance, $ \eta $, between $ T(\boldsymbol{x}) $ and the nearest inverted sphere $ T(W_j) $:
\begin{align} \label{eta distance}
\eta = \min_{1 \leq j \leq n+1} \text{dist} (T (\boldsymbol{x}), T(W_j) ) 
\end{align}
It is possible to write
\begin{align}
\text{dist} (T (\boldsymbol{x}), T(W_j) ) = ||T (\boldsymbol{x}) - \text{center}(T(W_j))|| - \text{radius}(T(W_j)) 
\end{align}
when $T(W_j)$ has finite radius. 
\begin{lemma}
	For the edges $ W_j $, $1 \leq j \leq n$,
	\begin{align} \label{center and radius simple edges}
	\text{center}(T(W_1)) & = (\alpha_1 - \frac{r^2}{2 \alpha_1}, \alpha_2,...,\alpha_n), \\
	\text{radius}(T(W_1)) & = \left| \frac{r^2}{2\alpha_1}  \right| = \frac{r^2}{2 \alpha_1},
	\end{align}
	and similar for the other $ W_j $, $ j \leq n $. For the face $ W_{n+1} $ defined by the equation $ x_1+ \cdots + x_n = 1 $, we get
	\begin{align} \label{center and radius Wn+1 edge}
	\text{center}(T(W_{n+1})) & = (\alpha_1 - \rho, ...,\alpha_n - \rho), \\
	\text{radius}(T(W_{n+1})) &= \sqrt{n}|\rho|, 
	\end{align}
	where
	\[ \rho = \frac{r^2}{2(\alpha_1 + \cdots + \alpha_n - 1)}. \]
\end{lemma}
\textit{Proof}: To derive (\ref{center and radius simple edges}), observe that since $ W_1 $ is given by the equation $ x_1 = 0 $, then $ T(W_1) $ is given by 
\begin{align}
T(W_1): \alpha_1 + \frac{r^2(x_1-\alpha_1)}{||\boldsymbol{x} - \boldsymbol{\alpha}||^2} = 0.
\end{align}
Completing the square in $ x_1 $ and letting $ \alpha_1 \neq 0 $, (or else $ T $ fixes $ W_1 $),

$$((x_1-\alpha_1)^2 + \cdots (x_n - \alpha_n)^2) + \frac{r^2}{\alpha_1}(x_1-\alpha_1)  = 0$$ \\

$$\left( x_1 - \left( \alpha_1 - \frac{r^2}{2 \alpha_1} \right) \right)^2 + (x_2 - \alpha_2)^2 + \cdots + (x_n - \alpha_n)^2  = \left( \frac{r^2}{2 \alpha_1} \right)^2 $$

By symmetry, the centers and radii for $ W_j $, $ j \leq n$ are similar. To derive (\ref{center and radius Wn+1 edge}), 
\begin{align}
T(W_{n+1}) : \alpha_1 + \frac{r^2(x_1-\alpha_1)}{||\boldsymbol{x} - \boldsymbol{\alpha}||^2} + \cdots + \alpha_n + \frac{r^2(x_n - \alpha_n)}{||\boldsymbol{x} - \boldsymbol{\alpha}||^2} = 1.
\end{align}
Letting $ \alpha_1 + \cdots + \alpha_n \neq 1 $, (or else $ T $ fixes $ W_{n+1} $) and completing the square now in each variable,
\begin{align*}
||\boldsymbol{x} - \boldsymbol{\alpha}||^2 + \frac{r^2(x_1+\cdots + x_n)}{(\alpha_1+ \cdots + \alpha_n - 1)} = \frac{r^2(\alpha_1+\cdots + \alpha_n)}{(\alpha_1+ \cdots + \alpha_n - 1)}.
\end{align*}
Letting $ \rho = \frac{r^2}{2(\alpha_1+ \cdots + \alpha_n - 1)} $,
\begin{align*}
& x_1^2 - 2x_1(\alpha_1 - \rho) + (\alpha_1 - \rho)^2\\
& + \\
& \vdots \\
& + x_n^2 - 2x_n(\alpha_n - \rho) + (\alpha_n - \rho)^2\\
& = 2 \rho (\alpha_1 + \cdots + \alpha_n) - \alpha_1^2 - \cdots - \alpha_n^2 + (\alpha_1 - \rho)^2 + \cdots + (\alpha_n - \rho)^2 \\
& = n \rho^2 = (\sqrt{n} \rho)^2 .
\end{align*}
$\square$

\noindent If a region of interest is bounded by the plane $ x_j = a $, then translation of (\ref{center and radius simple edges}) implies that 
\begin{align}
& \text{center} (T(\{ x_j  = a \})) = \left( \alpha_1,...,\alpha_j - \frac{r^2}{2(\alpha_j-a)},...,\alpha_n \right), \\
& \text{radius}(T(\{x_j  = a \}))  = \frac{r^2}{2|\alpha_j-a|}. 
\end{align}

\noindent An orthant of the unit ball in $ \mathbb{R}^n $ is comprised of the non-empty intersection of half spaces defined by $ W_1,...,W_n $ as well as the unit ball centered at the origin, $ B:x_1^2+ \cdots + x_n^2 = 1 $.

\begin{lemma}
	The $ T_{\boldsymbol{\alpha},r} $-image of $ B $ has center
	\begin{align}
	\left( \alpha_1 \left(1-\frac{r^2}{\lambda} \right),...,  \alpha_n \left( 1-\frac{r^2}{\lambda} \right) \right) \nonumber
	\end{align}
	and radius
	\begin{align}
	\frac{r^2}{|\lambda|} \nonumber
	\end{align}
	where $ \lambda = \alpha_1^2 + \cdots + \alpha_n^2 - 1 $.
\end{lemma}

\noindent \textit{Proof.} Let $ \dd = ||\bx - \baa||^2 \neq 0 $. $ T_{\boldsymbol{\alpha},r}(B) $ has equation
\[ \left( \alpha_1 + \frac{r^2(x_1-\alpha_1)}{\delta}  \right)^2 + \cdots + \left( \alpha_n + \frac{r^2(x_n-\alpha_n)}{\delta}  \right)^2 = 1. \]
Multiplying by $ \dd^2 $ and expanding terms,
\begin{align*}
\alpha_1^2\dd^2 + 2 \alpha_1 \dd r^2(x_1-\alpha_1) + r^4(x_1-\alpha_1)^2 + \cdots + \alpha_n^2\dd^2 + 2 \alpha_n \dd r^2(x_n-\alpha_n) + r^4(x_n-\alpha_n)^2 = \dd^2.
\end{align*}
Gathering $ r^4 \dd $ on the left hand side and dividing by $ \dd $,
\begin{align*}
\alpha_1^2 \dd + 2 \alpha_1 r^2(x_1-\alpha_1) + \cdots + \alpha_n^2 \dd + 2 \alpha_n r^2(x_n-\alpha_n) + r^4 = \dd.
\end{align*}
Letting $ \lambda = \alpha_1^2 + \cdots + \alpha_n^2 - 1 \neq 0 $, we get
\begin{align*}
\dd \lambda + 2\alpha_1r^2x_1 + \cdots + 2\alpha_n r^2 x_n = 2\alpha_1^2 + \cdots + 2\alpha_n^2r^2 - r^4.
\end{align*}
Letting $ \sigma_j = \frac{2\alpha_jr^2}{\lambda} $, 
\begin{align*}
\dd + \sigma_1x_1 + \cdots + \sigma_n x_n = \frac{1}{\lambda}(2r^2(\alpha_1^2 + \cdots +\alpha_n^2)-r^4) = \tau. 
\end{align*}
\begin{align*}
& x_1^2 - 2x_1 \left( \alpha_1 - \frac{\sigma_1}{2} \right) + \left( \alpha_1 - \frac{\sigma_1}{2} \right)^2 + \\
& \vdots \\
& x_n^2 - 2x_n \left( \alpha_n - \frac{\sigma_n}{2} \right) + \left( \alpha_n - \frac{\sigma_n}{2} \right)^2 \\
& = \tau - (\alpha_1^2+ \cdots + \alpha_n^2) + \left(\alpha_1 - \frac{\sigma_1}{2} \right)^2 + \cdots + \left(\alpha_n - \frac{\sigma_n}{2} \right)^2.
\end{align*}
The left hand side consists of perfect squares as desired and the right hand is
\begin{align*}
& \frac{1}{\lambda}(2r^2(\alpha_1^2 + \cdots+ \alpha_n^2)-r^4) - \alpha_1^2 - \cdots - \alpha_n^2 \\
& + \alpha_1^2 - \frac{2\alpha_1^2r^2}{\lambda} + \frac{\alpha_1^2 r^4}{\lambda^2} + \cdots + \alpha_n^2 - \frac{2 \alpha_n^2 r^2}{\lambda} + \frac{\alpha_n^2 r^4}{\lambda^2} \\
& = \frac{\alpha_1^2r^4}{\lambda^2} + \cdots + \frac{\alpha_n^2r^4}{\lambda^2} - \frac{r^4 \lambda}{\lambda^2} = \frac{r^4}{\lambda^2}.
\end{align*}
$ \square $

\section{Inversion Properties}
\label{sec:app-B}
The Jacobian for $ T_{\boldsymbol{\alpha},r} $ is given by
\begin{align}
J = \begin{pmatrix}
\frac{\partial u_1}{\partial x_1} & \frac{\partial u_1}{\partial x_2} & \cdots & \frac{\partial u_1}{\partial x_n} \\
\frac{\partial u_2}{\partial x_1} & \frac{\partial u_2}{\partial x_2} & \cdots & \frac{\partial u_2}{\partial x_n} \\
\vdots & \vdots & \ddots & \vdots \\
\frac{\partial u_n}{\partial x_1} & \cdots & \cdots & \frac{\partial u_n}{\partial x_n}
\end{pmatrix}
\end{align}
where $ u_j = \alpha_j + 2 \rho (x_j - \alpha_j) $, the $ j^{\text{th}} $ component of $ T_{\boldsymbol{\alpha},r}(\boldsymbol{x}) $. Taking partial derivatives, we get
\[ [J]_{i,j} = \begin{cases}
\frac{r^2(||\boldsymbol{x} - \boldsymbol{\alpha}||^2 - 2(x_j - \alpha_j)^2)}{||\boldsymbol{x} - \boldsymbol{\alpha}||^4}, & \quad i = j \\
\frac{-2r^2(x_i-\alpha_i)(x_j - \alpha_j)}{||\boldsymbol{x} - \boldsymbol{\alpha}||^4}, & \quad i \neq j\\
\end{cases} \]
Pulling out the factor of $ \frac{r^2}{||\boldsymbol{x} - \boldsymbol{\alpha}||^4} $ and making the substitution $ a_j = (x_j - \alpha_j) $, we can realize
\begin{align*}
[J]_{i,j} = \frac{r^2}{\left( \sum_{k=1}^n a_k^2 \right)^2} 
\begin{cases}
\sum_{k=1}^n a_k^2 - 2a_j^2, & \quad i = j \\
-2a_ia_j, & \quad i \neq j\\
\end{cases}
\end{align*}
from which the binomial theorem and cofactor expansion gives 
\[ \det J = \left( \frac{r^2}{\left( \sum_{k=1}^n a_k^2 \right)^2} \right)^n (-(\sum_{k=1}^\infty a_k^2)^n) = - \left( \frac{r}{||\boldsymbol{x} - \boldsymbol{\alpha}||}  \right)^{2n} \]

\begin{figure}[h]
	\caption{Projection $P: \bx \mapsto \boldsymbol{\alpha} $ induces a partition of $ \mathcal{S}_- $. The image of one region in the two dimensional case is shown. It is the intersection of two hyperbolas and is contained in $ [2+\sqrt{2},\infty) \times [0,1] $ if $ r=1 $ or $ [2(2+\sqrt{2}),\infty) $ if $ r = \sqrt{2} $. \label{pic: 3 regions 2d simplex A image}}
	\label{fig:Ppartition}
	\begin{center}
		\includegraphics[scale=0.2]{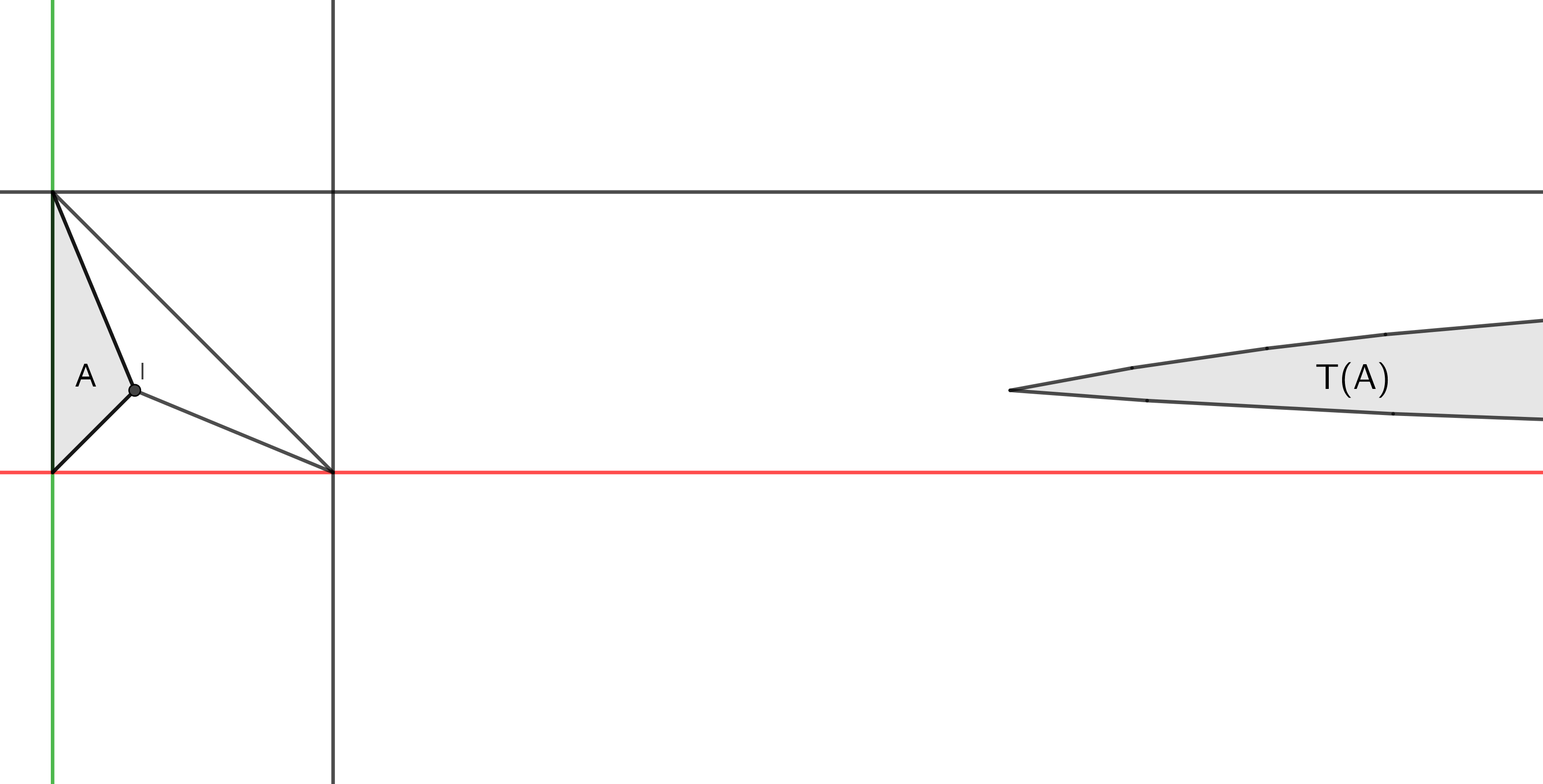}
	\end{center}
\end{figure}

\noindent Let $ P(\bx) =  \underset{\boldsymbol{y}\ \in\ \partial(\mathcal{S}_-)}{\mathrm{argmin}}\ ||\boldsymbol{x} - \boldsymbol{y} || $ be projection of $ \bx $ onto the nearest face $ W_j $. Note that $P$ partitions $ \mathcal{S}_- $ into disjoint sets. For example, when $ \mathcal{S}_- $ is a two-dimensional isosceles triangle, $P$ partitions $ \mathcal{S}_- $ into four regions; three triangles which project to a common face and a set of three line segments containing the incenter, $ I $, of planar Lebesgue measure zero as shown in Figure \ref{fig:Ppartition}. 

\begin{lemma}
	$ T_{P(\cdot),\sqrt{2}} $ is an injective function on $ \mathcal{S}_- $. 
\end{lemma}

\noindent \textit{Proof.}
First, we establish that $ T_{P(\cdot),\sqrt{2}} $ is well-defined. The set $\mathcal{Q} =  \{ \bx: \text{dist}(\bx,W_j) = \text{dist}(\bx,W_i), i \neq j \} $ has Lebesgue measure zero. In case $ \bx \in \mathcal{Q} $, define $ P(\bx) $ to project onto $ j $ for $ j>i $. This defines a unique projection onto a face for all $ \bx  \in \mathcal{S}_-$ making $ P $ well-defined. Since $ T $ is inversion in a sphere, it too is well-defined and thus $ T_{P(\cdot),\sqrt{2}} $ is well-defined. There are now three cases to consider:

\ti{Case 1}: $ P(\bx) = P(\by) $. Then, if $ T_{P(\bx),\sqrt{2}}(\bx) = T_{P(\by),\sqrt{2}}(\by) $ then $ \bx = \by $ since inversion is injective.

\ti{Case 2}: $ P(\bx),P(\by) \in{W_j} $ but $ P(\bx) \neq P(\by) $ (in which case $ \bx \neq \by $). Then, $ T_{P(\bx),\sqrt{2}}(\bx) $ lies in a line perpendicular to $ W_j $ through $ P(\bx) $ and $ T_{P(\by),\sqrt{2}}(\by) $ lies in a line perpendicular to $ W_j $ through $ P(\by) $. Since $ P(\bx) \neq P(\by) $, these lines are disjoint, thus $ T_{P(\bx),\sqrt{2}}(\bx) \neq T_{P(\by),\sqrt{2}}(\by) $.

\ti{Case 3}: $ P(\bx) \in{W_j} $ and $ P(\by) \in{W_i} $, $ i \neq j $ (and again $ \bx \neq \by $ by definition of $ P $). Since $ \mathcal{S}_- $ is contained in the hypercube $ [0,1]^n $, then for $ j \leq N $, $ W_j $ is contained in a face of $ [0,1]^n $. Since the radius of inversion is fixed at $ r = \sqrt{2} $, then observe that
\begin{align*}
T(W_1)  & \subset [2,\infty) \times [0,1]^{n-1} \\
& \vdots \\
T(W_j) & \subset [0,1]^{j-1} \times [2,\infty) \times [0,1]^{n-j}\\
& \vdots \\
T(W_n) & \subset [0,1]^{n-1} \times [2,\infty) \\
T(W_{n+1}) & \subset (- \infty ,0]^n
\end{align*} 
comprising $ n+1 $ disjoint sets. Thus, if $ \bx $ and $ \by $ are projected onto different faces, then $ T_{P(\bx),\sqrt{2}}(\bx) \neq T_{P(\by),\sqrt{2}}(\by) $. $ \square $

\noindent The details showing that $ T_{P(\cdot),\sqrt{2}} $ is injective on a sector of an $ n $-sphere as well as the $ n $-dimensional hypercube follow similarly.

\newpage
\bibliographystyle{plain}

\bibliography{SIS}

\thanks{\nd \small Contact: sharang.chaudhry@gmail.com, daniel.lautzenheiser@gmail.com, kaushik.ghosh@unlv.edu}

\end{document}